\documentclass[twocolumn]{aastex631}
\usepackage{amsmath}
\let\tablenum\relax 
\usepackage{siunitx}
\usepackage{graphicx,subfigure,float}
\usepackage{gensymb}
\usepackage{booktabs}

\begin{document}

\title{Time-variable Scattered Light in Herbig Disks Observed with Subaru/SCExAO\footnote{This research is based in part on data collected at the Subaru Telescope, which is operated by the National Astronomical Observatory of Japan.}}

\author[0009-0007-3210-4356]{Camryn Mullin}
\affiliation{Department of Physics and Astronomy, University of Victoria, Victoria, BC, V8P 5C2, Canada}

\author[0000-0001-6341-310X]{Miles Lucas}
\affiliation{Steward Observatory, Department of Astronomy, The University of Arizona, Tucson, AZ 85721, USA}

\author[0000-0001-9290-7846]{Ruobing Dong}
\affiliation{Kavli Institute for Astronomy and Astrophysics, Peking University, Beijing 100871, China}
\affiliation{Department of Physics and Astronomy, University of Victoria, Victoria, BC, V8P 5C2, Canada}

\author[0000-0002-3053-3575]{Jun Hashimoto}
\affil{Academia Sinica Institute of Astronomy \& Astrophysics (ASIAA), 11F of Astronomy-Mathematics Building, AS/NTU, No.1, Sec. 4, Roosevelt Rd., Taipei 106319, R.O.C.}
\affiliation{Astrobiology Center, National Institutes of Natural Sciences, 2-21-1 Osawa, Mitaka, Tokyo 181-8588, Japan}

\author[0000-0003-2948-5614]{Haochang Jiang}
\affiliation{Max-Planck-Institut für Astronomie, Königstuhl 17, 69117 Heidelberg, Germany}

\author[0000-0002-6773-459X]{Doug Johnstone}
\affiliation{NRC Herzberg Astronomy and Astrophysics, 5071 West Saanich Rd, Victoria, BC V9E 2E7, Canada}
\affiliation{Department of Physics and Astronomy, University of Victoria, Victoria, BC, V8P 5C2, Canada}

\author[0000-0002-6964-8732]{Kellen Lawson}
\affiliation{Center for Space Sciences and Technology, University of Maryland, Baltimore County, Baltimore, MD 21250, USA}
\affiliation{Astrophysics Science Division, NASA-GSFC, Greenbelt, MD 20771, USA}
\affiliation{Center for Research and Exploration in Space Science and Technology, NASA-GSFC, Greenbelt, MD 20771, USA}

\author[0000-0001-5638-1330]{Sean Brittain}
\affiliation{Department of Physics and Astronomy, Clemson University, Clemson, SC 29634-0978, USA}

\author[0000-0002-1097-9908]{Olivier Guyon}
\affiliation{Subaru Telescope, National Astronomical Observatory of Japan, 650 North A‘ohōkū Place, Hilo, HI 96720, USA}
\affiliation{Steward Observatory, Department of Astronomy, The University of Arizona, Tucson, AZ 85721, USA}
\affiliation{College of Optical Sciences, University of Arizona, Tucson, AZ 85721, USA}
\affiliation{Astrobiology Center, National Institutes of Natural Sciences, 2-21-1 Osawa, Mitaka, Tokyo 181-8588, Japan}

\author{Tomoyuki Kudo}
\affiliation{Subaru Telescope, National Astronomical Observatory of Japan, 650 North A‘ohōkū Place, Hilo, HI 96720, USA}

\author[0000-0002-3047-1845]{Julien Lozi}
\affiliation{Subaru Telescope, National Astronomical Observatory of Japan, 650 North A‘ohōkū Place, Hilo, HI 96720, USA}

\author[0000-0002-5758-150X]{Joan Najita}
\affiliation{NOIRLab, 950 North Cherry Avenue, Tucson, AZ 85719, USA}

\author[0000-0003-1526-6787]{He Sun}
\affiliation{Academy for Advanced Interdisciplinary Studies, Peking University, Beijing 100871, China}
\affiliation{College of Future Technology, Peking University, Beijing 100871, China}

\author[0000-0002-6510-0681]{Motohide Tamura}
\affiliation{Astrobiology Center, National Institutes of Natural Sciences, 2-21-1 Osawa, Mitaka, Tokyo 181-8588, Japan}
\affiliation{National Astronomical Observatory of Japan, 2-21-2, Osawa, Mitaka, Tokyo 181-8588, Japan}
\affiliation{Department of Astronomy, Graduate School of Science, The University of Tokyo, 7-3-1, Hongo, Bunkyo-ku, Tokyo 113-0033, Japan}

\author[0000-0002-4309-6343]{Kevin Wagner}
\affiliation{Steward Observatory, Department of Astronomy, The University of Arizona, Tucson, AZ 85721, USA}

\correspondingauthor{Camryn Mullin, Ruobing Dong}
\email{camrynmullin@uvic.ca, rbdong@pku.edu.cn}

\begin{abstract}
Using the Subaru Coronagraphic Extreme Adaptive Optics (SCExAO) instrument, we present near-infrared $K$-band polarimetric imaging of nine Herbig stars selected from a volume-limited sample within 200~pc. We detect the disks around MWC~480, HD~163296, and HD~143006 for the first time with SCExAO, and compare these observations with previous VLT/SPHERE datasets to identify surface-brightness variability.
In MWC~480, we resolve two azimuthal brightness dips near the disk minor axis and find evidence that one of them shifted between 2021 and 2022.
In HD~163296, we identify an apparent linear azimuthal motion of a localized peak in polarized intensity along the outer ring over a 15-month baseline. The rapid motion of these features relative to the local Keplerian velocity suggests that the observed variability is driven by changing illumination rather than physical material motion. Due to uncertainties in the underlying scattering background, however, we cannot determine the precise physical origin of the variability. No significant disk variability is detected in HD~143006 over a 10-month baseline.
We also report the first detection of a protoplanetary disk using the fast-PDI mode on SCExAO, illustrating both the promise and current limitations of this observing mode. Finally, we report non-detections toward HD~144432, HD~56895, PDS~76, HIP~80425, HD~148352, and HIP~81474. All non-detections with Meeus classifications belong to Group~II systems and are likely self-shadowed. For these six systems, we measure the system-integrated polarization fraction and angle of linear polarization, providing quantitative constraints on their unresolved circumstellar environments.
\end{abstract}


\section{Introduction} \label{sec:intro}
Protoplanetary disks have been found to have complex morphological features such as rings and gaps, spiral arms, large cavities, and kinematic structures \citep{bae23, benisty23, pinte23}. Theories suggest that these features may be caused by interactions between forming planets and their host disk (e.g., \citealt{dong15spiralarm,dong15gap,Bae2017, Bae2018}). Recent advances in observational capabilities are beginning to uncover connections between disk substructures and planet formation. For example, \citet{Wagner2019} and \citet{Currie2022} each reporting a planet candidate associated with a spiral arm in MWC 758 and AB Aurigae respectively, \citet{Hammond2023} reporting a planet candidate in a disk gap of HD 169142, recent papers by \cite{vanCapelleveen2025} and \cite{close25} detecting the gap opening planet WISPIT2b, and the detections of the confirmed planets in the cavity of PDS 70 \citep{Keppler2018, wagner18pds70, haffert19}. There have also been multiple planet candidates suggested through kinematic ``kinks" or ``Doppler flip" in disk gas structure \citep{Pinte2018,Pinte2019, casassus19}.

These dust and gas substructures offer valuable insight into planet formation environments, but the limited angular resolution of our current scattered-light instruments only allow detection of structures and planets separated by tens of astronomical units (au). These inner au can be probed with optical and near-infrared (NIR) interferometry to reveal inner disk misalignments, \citep{Dullemond2010, Bohn2022}, however interferometric observations have limited sensitivity and often require modeling to produce image results. Therefore, our understanding of inner disks from interferometry is complemented by scattered light observations, which reveal an indirect method to constrain inner disk dynamics through shadows cast on the resolvable outer disk. These shadows can be cast by either a misaligned or puffed-up inner disk, or a planet -- in principle, any structure that blocks starlight from the observable disk \citep{benisty23}. The morphology of a shadow can help determine its source. A pair of narrow shadows suggest a highly misaligned inner disk (e.g., \citealt{marino15hd142527,Benisty2017,Hunziker2021}), while a broad shadow (sometimes covering up to half the disk, \citealt{Debes2017, Benisty2018, Muro2020, Hashimoto2024}) indicates a small misalignment or a warp in the inner disk. 

Shadowing can directly affect disk dynamics and planet formation within the disk. Due to the cooling of the shadowed material, a pressure imbalance is created which can lead to the formation of substructures. For example, shadows can trigger the formation of spiral arms \citep{Montesinos2018, Cuello2019, muley25, Ziampras2025}, and warps \citep{zhang25}.

Some shadows show azimuthal movement, changes in width, or disappearances/appearances on timescales of weeks to years \citep{Debes2017,Stolker2017,Pinilla2018, Bertrang2020, debes2023, Lucas2025}. Moving shadows can be explained by procession of the inner disk (potentially due to torquing by an inner planet), by changes in inner disk scale height, or by an orbiting inner companion \citep{benisty23, Akansoy2025}. The timescale of shadow movement can help determine the cause of change in a shadow. 

\begin{table*}[!htpb] 
\centering
\caption{Target Properties}
\label{tab:observations}
\hspace*{-3cm}\begin{tabular}{lccccccccc}\hline
\
Target & RA\tablenotemark{a} & Dec\tablenotemark{a} & Distance\tablenotemark{h} & Spectral & Meeus  & $M_\star$ & Log$(L_\star)$  & $M_{\rm disk}$\tablenotemark{a} 
& SCExAO Disk \\ 
 & (h:m:s) &  (d:m:s) & (pc) & Type & Group & ($M_\odot$) & $(L_\odot)$ & $(M_\odot)$ 
 & Detected? \\ \hline
MWC 480 & 04:58:46.3 & +29:50:37 & $156\pm1$& A5-A6\tablenotemark{a} & II\tablenotemark{a} & 2.1\tablenotemark{f} & $2.05^{+0.01}_{-0.01}$\tablenotemark{a} & $0.051\pm4\rm{E}$-4 & Yes \\
HD 144432 & 05:58:36.9 & -22:57:16& $154.8\pm0.5$& G7-G9\tablenotemark{a} & II\tablenotemark{a} & 1.82\tablenotemark{i} & $0.54^{+0.02}_{-0.02}$\tablenotemark{a} & $0.021\pm0.002$ & No \\ 
HD 56895 & 07:18:31.8 & -11:11:34 & $165.8\pm0.6$& F2-F3\tablenotemark{a} & II\tablenotemark{a}& 1.64\tablenotemark{a} & $0.98^{+0.01}_{-0.01}$\tablenotemark{a} & - & No \\
PDS 76 & 15:56:40.0 & -22:01:40 & $146.4\pm0.4$& A0-F9\tablenotemark{a} & II\tablenotemark{a} & 1.78\tablenotemark{g} & $1.13^{+0.01}_{-0.01}$\tablenotemark{a} & $0.013\pm0.001$& No   \\
HD 143006 & 15:58:36.9 & -22:57:16 & $167.3\pm0.5$& G7-G9\tablenotemark{a} & I\tablenotemark{a}& 1.4\tablenotemark{f} & $0.54^{+0.02}_{-0.02}$\tablenotemark{a} & $0.032\pm5\rm{E}$-5  & Yes \\
HIP 80425 & 16:24:59.1 & -25:21:18 & $135.7\pm0.6$& A1\tablenotemark{d}& - & 2.39\tablenotemark{g} & 1.20\tablenotemark{c} & - & No \\
HD 148352 & 16:28:25.2  & -24:45:01 & $76.4\pm0.2$& F2\tablenotemark{e} & - &1.9\tablenotemark{g}& 1.21\tablenotemark{e} & - & No \\
HIP 81474 & 16:38:28.6 & -18:13:14 & $154.3\pm0.7$ & B9-A0\tablenotemark{a}& II\tablenotemark{a}& 2.73\tablenotemark{g} & $1.22^{+0.01}_{-0.01}$\tablenotemark{a} & - & No  \\ 
HD 163296 & 17:56:21.3 & -21:57:21 & $101.0\pm0.4$& A2-A3\tablenotemark{a}& I\tablenotemark{b} & 2.0\tablenotemark{f} & $1.19^{+0.04}_{-0.05}$\tablenotemark{a} & $0.034\pm0.001$& Yes     \\
\hline
\end{tabular}
\tablenotetext{a}{\cite{Guzman2021}; 
                  \tablenotemark{b} \cite{Muro2018}; 
                  \tablenotemark{c} \cite{Dong2018};
                  \tablenotemark{d} \cite{Chen2012};
                  \tablenotemark{e} \cite{Erickson2011};
                  \tablenotemark{f} \cite{Ren2023};
                  \tablenotemark{g} \cite{Pecaut2013};
                  \tablenotemark{h} \cite{Gaia2023};
                  \tablenotemark{i} \cite{Muller2011}}

\end{table*}

In addition to shadowing, some disks show variability in brightness peaks. The $\sim60$au ring of HD 163296 has shown azimuthal brightness variation between observational epochs \citep{Rich2019, Muro2018}. Such variation could be attributed to small changes in scale height, changes in local dust distribution, time variable stellar accretion, or potential inner disk misalignment causing weak shadows \citep{Rich2019}. \cite{Ma2024} found an increase in polarimetric brightness in PDS 70, which could be attributed to reduced shadowing from an evolving inner disk, or variable stellar illumination due to variable accretion.

In this work, we aim to study the disks around young intermediate-mass stars and their link to planet formation. The protoplanetary disks around such stars (often called Herbig disks) are of particular interest to planet formation studies.  Main sequence intermediate-mass stars are favored for hosting large planets, with giant planet demographics peaking for A-type stars \citep{Wolthoff2022, Brittain2023}. Since young intermediate mass (or Herbig) stars are mainly the precursors for A and B type stars, their disks are also more likely to possess young giant planets than disks around T-Tauri stars. Many Herbig stars possess (on average) larger disks in both radius and mass than disk around T-Tauri stars, with structures key for planet formation statistics \citep{Stapper2022, stapper2024, Stapper2025}. Some of these structures, such as two-arm spirals, are best explained by giant planet formation \citep{Ren2018, ren20}. 

This paper is organized as follows: \S\ref{sec:methods} covers our targets, observational setup, and data reduction, \S\ref{sec: analysis} compares our CHARIS data to past observations with VLT/SPHERE/IRDIS, in \S\ref{sec:Discussion} we discusses our results, and in \S\ref{sec:conclusions} we present our conclusions.

\begin{figure*}[!htpb]%
    \centering
    \includegraphics[width=0.9\textwidth]{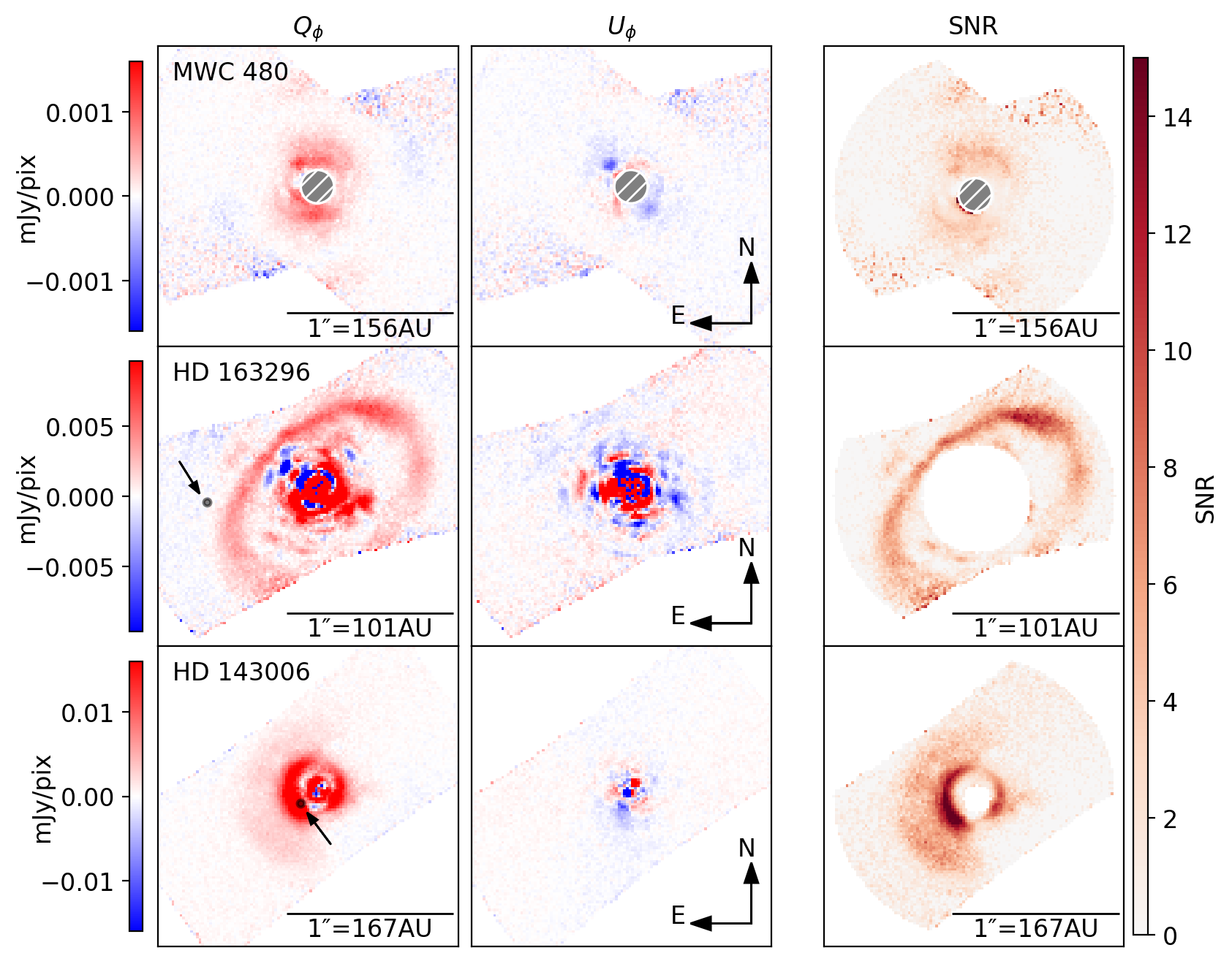}
    \caption{CHARIS PDI results showing $Q_\phi$ \textbf{(left)} and $U_\phi$ \textbf{(middle)} for the 3 previously-detected disks. The color bar is set such that red is positive and blue is negative. Each disk is detected as can be seen by the positive $Q_\phi$ and noise-like $U_\phi$. The 0\farcs1 coronagraphic IWA for MWC 480 is marked by the hatched region. The black points (indicated with arrows) in the left column show the approximate location of the kinematically predicted planets from \cite{Pinte2020} assuming CCW rotation (same as the disk; \citealt{Barenfeld2016, Teague2019}) at Keplerian speed over a 5 year baseline. 
    The right column shows radial SNR. Signal data was extracted from $Q_\phi$ and radial noise from $U_\phi$ using equation \ref{eq:mad std}. The central star is masked out and does not contribute to the calculation. (The $Q_\phi$ and $U_\phi$ fits files are available in the online version. DOI: \url{https://doi.org/10.3847/1538-3881/ae473d})} 
    \label{fig:previously observed}
\end{figure*}

\section{Observations and Data Reduction} \label{sec:methods}
This section describes how the data were obtained and presents the results of the observations. In section \ref{subsec: Sample} we present our observational sample. In section \ref{subsec: observations} we explain the general observational set-up for our nine targets. In section \ref{subsec: data reduciton} we outline how our PDI data products were obtained. In section \ref{subsec:charis} we discuss the CHARIS-specific data reduction and present the CHARIS results. In section \ref{subsec:fast-pdi} we do the same for fast-PDI. 

\begin{figure}[!htpb]%
    \centering
        {\includegraphics[width=0.45\textwidth]{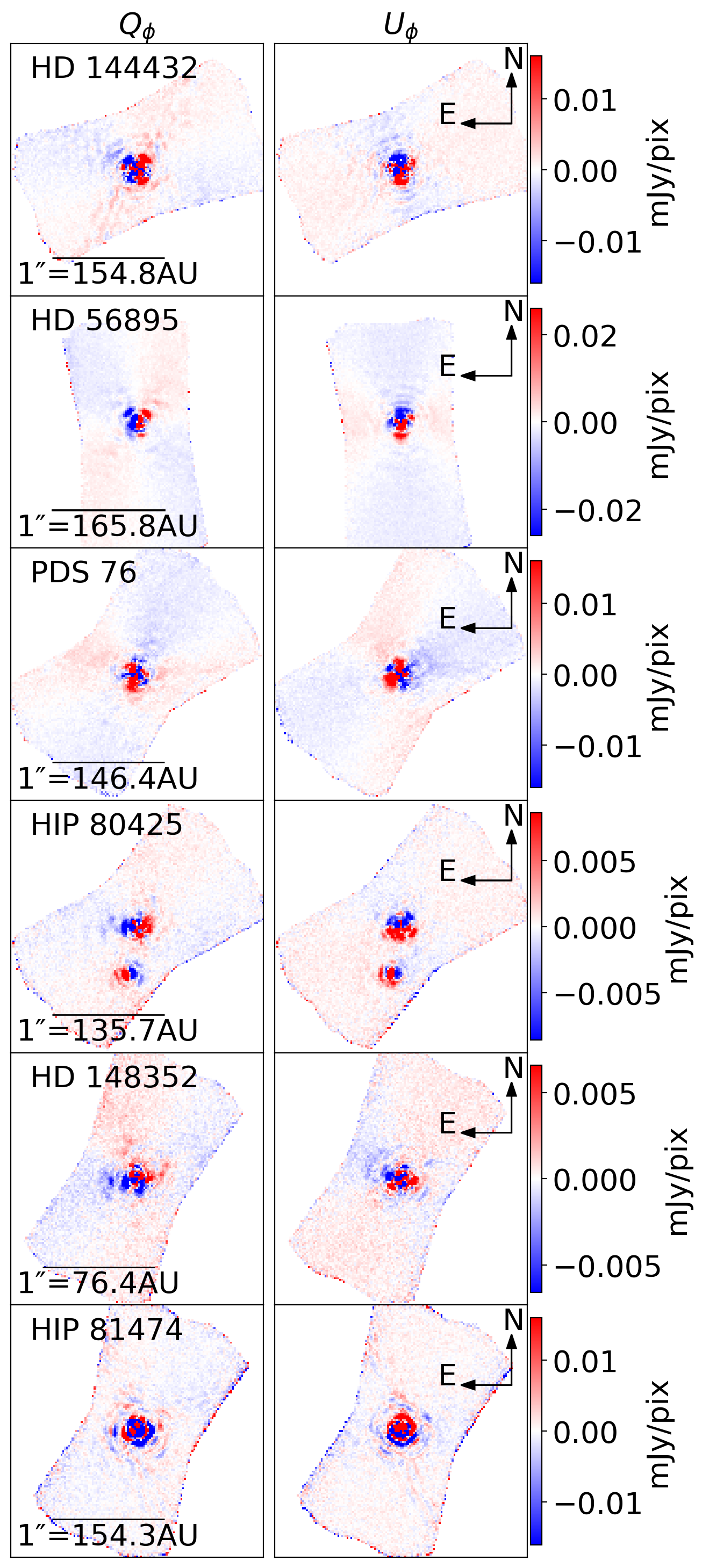}}
    \caption{CHARIS PDI results showing azimuthal Stokes for the six previously unobserved targets. The quadrupole pattern in $Q_\phi$ is consistent with non-detection of extended sources (disk) and polarized point-source (star) detection. (The data are available in the online version.)}
    \label{fig:CHARIS_PDI}
\end{figure}

\subsection{Sample} \label{subsec: Sample}
Our targets were selected from a volume-limited sample of 21 Herbig stars within 200 pc \citep{Dong2018}\footnote{The master 200 pc Herbig sample originally contains 24 objects \citep{Dong2018}. Three of them have had their distances revised to be outside 200 pc by \citet{Gaia2023}. All nine targets studied in this paper are still within 200 pc and are therefore part of the volume-limited sample.}, with the main goal of observing some of the remaining unimaged disks accessible using Subaru.
The master sample includes only Herbig sources that have spectral type F6 or earlier, $M_\star>1.5M_\odot$, and no known stellar companion within 5\arcsec. This totals 21 known Herbig stars, some of which possess well-studied disks such as AB Aurigae, MWC 758, SAO 206462 and HD 100546, while others do not have high resolution infrared disk detections. At the time of observing, twelve of the 21 targets had been imaged previously in IR, leaving nine remaining. Six of these nine --- HD 56895, HD 144432, HIP 80425 (HD 147807), HIP 81474 (HD 149914), HD 148352, and PDS 76 (HD 142666) --- are visible from Hawaii. We have observed them for the first time with the Subaru Coronagraphic Extreme Adaptive Optics instrument (SCExAO, \citealt{Jovanovic2015}). SCExAO has multiple optical paths designed to detect disks around young stars which is why this instrument was selected for this disk study. 

Additionally, we used SCExAO to observe three more targets selected from the sample of 21: HD 163296, HD 143006 and MWC 480 (HD 31648). These targets have disk structures that have been previously resolved in the NIR with instruments such as the Very Large Telescope (VLT) Spectro-Polarimetric High-contrast Exoplanet REsearch (SPHERE, \citealt{Beuzit2019}) instrument, the Subaru High-Contrast Coronographic Imager for Adaptive Optics (HiCIAO, \citealt{Hodapp2008}) camera, and the Gemini Planet Imager (GPI, \citealt{Macintosh2008}). Observing these three targets with SCExAO serves the dual purpose of allowing us to study variability in Herbig disks by comparing our SCExAO epochs with similar imaging, and providing a base-line for what disk structures we can hope to resolve with SCExAO to better understand any new disk detections. 

\begin{figure}[!htpb]%
    \centering
    {\includegraphics[width=0.4\textwidth]{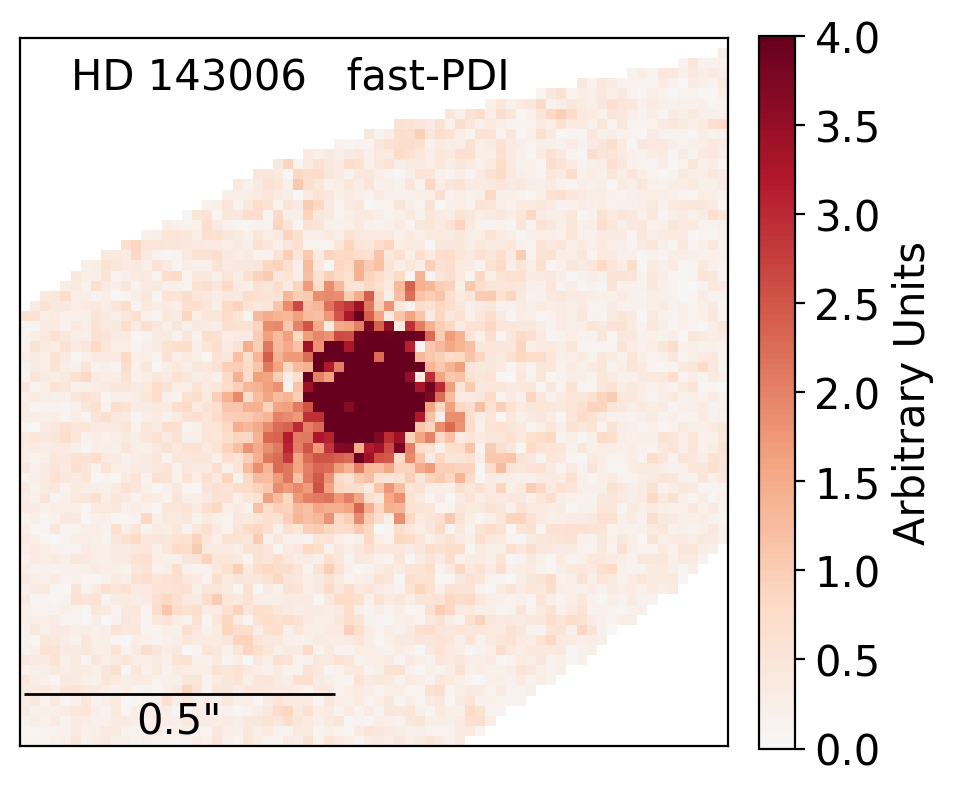}}
    \caption{
    The fast-PDI results for HD 143006. The units are normalized intensity, with the color-bar truncated to allow the disk to be seen. The inner IR ring is visible on the edges of the PSF but the outer ring is obscured by background scatter. (The data are available in the online version.)}
    \label{fig:fast_PDI}
\end{figure}

\subsection{Observations} \label{subsec: observations}
We obtained data with SCExAO in two configurations: the Coronagraphic High Angular Resolution Imaging Spectrograph (CHARIS, \citealt{Chilcote2019}) operated in polarimetric differential imaging mode (PDI, \citealt{Joost2021, Lawson2021}), and the fast near-infrared polarimetric differential imaging (fast-PDI, \citealt{Lozi2020}) mode. PDI is advantageous for detecting disks in IR scattered light since it allows for removal of un-polarized starlight while retaining polarized light scattered by disk dust grains \citep{Kuhn2001}. CHARIS has successfully imaged the Herbig disk AB Aurigae and proven to be ideal for disk detection \citep{Currie2022}. 

The six previously unimaged targets were observed as part of the same program (PI: R. Dong) over three nights. HD 56895 was observed on 2022 January 10, and the other five over two half nights, 2022 March 27 and 28. All observations were carried out in PDI mode, non-coronagraphic, $K$-band ($2.01–2.36 \mu$m), with simultaneous fast-PDI $H$-band ($1.47–1.79 \mu$m) observations. 

The additional three targets were observed as part of separate programs, and added to this analysis as a Herbig disk comparison sample. HD 163296 and HD 143006 (PI: J. Hashimoto) were observed on 2022 April 8 using the same observing set-up as the six targets above. MWC 480 (PI: H. Jiang) was observed on 2022 November 24, and is the only coronagraphic observation in the dataset. Specifically, the Lyot coronagraph with a 113 mas occulter. The MWC 480 observations were broadband, however, we opted to only use the $K$-band portion of the data for consistency with the other datasets. The observational results for each observing mode will be discussed in more detail in sections \ref{subsec:charis} and \ref{subsec:fast-pdi}.

\subsection{Data Reduction Methods} \label{subsec: data reduciton}

We used the double differencing technique described in \cite{Hinkley2009} and in the appendix of \cite{Oppenheimer2008} to subtract the stellar PSF. The angle of polarization in the SCExAO optical path is determined by the half-wave-plate (HWP), which filters for polarized light at 4 rotator angles (\ang{0}, \ang{45}, \ang{22.5} and \ang{67.5}, relative to the instrument optical bench) completing a full HWP cycle with minimal parallactic rotation between exposures. The HWP filtered light is then sent to the Wollaston beam-splitter, which shifts the phase of part of the incoming beam such that two separate phases of light exit the beam splitter and hit the left and right sides of the detector orthogonally. At a HWP angle of \ang{0} this corresponds to Stokes ``$Q$" vector components $Q^+_{L}$ and $Q^+_{R}$. These orthogonal ``left" and ``right" images are combined to create the Stokes $Q$ vector using two sets of HWP angles each. At \ang{0} and \ang{45} $Q^+$ and $Q^-$ are calculated respectively from the left and right images,
\begin{align}
    Q^+_{R} - Q^+_{L} = Q^+ \label{eq:stokes first} \\ 
    Q^-_{R} - Q^-_{L} = Q^- 
\end{align}
and then subtracted again to obtain the $Q$ image,
\begin{equation}
    \frac{1}{2}(Q^+ - Q^-) = Q \label{eq:stokes}
\end{equation} 
and likewise for Stokes $U$ at \ang{22.5} ($U^+$) and \ang{67.5} ($U^-$). Once the full cycles are obtained, the polarized intensity ($PI$) signal is then calculated as $PI = \sqrt{Q^2+U^2} $. Since the squaring of $Q$ and $U$ can lead to an increase in photon noise, a better measure of signal and noise is estimated by the azimuthal Stokes vectors $Q_\phi$ and $U_\phi$ such that,
\begin{align} 
    Q_\phi = -Q\cos(2\phi) - U \sin(2\phi)  \\
    U_\phi = +Q\sin(2\phi) - U \cos(2\phi)\label{eq:uphi}
\end{align}
\citep{Schmid2006,Avenhaus2014, Monnier2019} where $\phi$ is the position angle (measured from North-up) relative to the center of the image (star's location). This convention assumes polarization ``signal" is azimuthal and thus positive in $Q_\phi$, which holds true for disks where polarization is the result of single scattering events, and the disk is assumed to have minimal inclination. $U_\phi$ takes polarization \ang{45} offset with respect to the azimuth and thus contains (ideally) no signal unless the disk is highly inclined \citep{Schmid2021}. An inclined disk will experience multiple scattering along the line of sight such that $U_\phi$ can no longer be considered as simply noise \citep{Canovas2015}.

\begin{table*}[!htpb] 
\centering
\caption{Observations of Detected Disks}
\label{tab:obs info}
\hspace*{-3cm}\begin{tabular}{lccccccccccc}\hline
\
&Target & Date & Telescope & Instrument & Filter & IWA & pix scale & FWHM & DIT & $T_{exp}$ & Reference \\ 
 & & (UTC) &  &  & & (mas) & (mas/pix) & (pix) & (s) & (min) & \\ \hline
1. &MWC 480 & 2010/01/24 & Subaru & HiCIAO & H & 150 & 9.50 & 5 & 19.5 & - & [a]\\ 
2. &MWC 480 & 2018/11/27 & VLT & IRDIS & H & $\sim$50\tablenotemark{d} & 12.25 & 4 & 32.0 & 57.0 & [c] \\
3. &MWC 480 & 2021/12/10 & VLT & IRDIS & K & 100 & 12.25 & 5 & 16.0 & 51.0& [b] \\
4. &MWC 480 & 2022/11/24 & Subaru & CHARIS & K & 113 & 16.20 & 4 & 60.5 & 256.0 & - \\
\hline
5. &HD 163296 & 2021/04/06 & VLT & IRDIS & K & 100 & 12.25 & 5 & 16.0 & 6.4 & [b]\\
6. &HD 163296 & 2021/06/03 & VLT & IRDIS & K & 100 & 12.25 & 5 &16.0 & 34.0 & [b]\\
7. &HD 163296 & 2021/09/09 & VLT & IRDIS & K & 100 & 12.25 & 5 &16.0 & 25.6 & [b]\\
8. &HD 163296 & 2021/09/26 & VLT & IRDIS & K & 100 & 12.25 & 5 & 16.0 & 8.5 & [b]\\
9. &HD 163296 & 2022/04/08 & Subaru & CHARIS & K & sat (326) & 16.20 & 4 & 31.0 & 60.0 & - \\
10. &HD 163296 & 2022/06/11 & VLT & IRDIS & K & 100 & 12.25 & 5 & 64.0 & 51.0 & [b]\\
11. &HD 163296 & 2022/07/07 & VLT & IRDIS & K & 100 & 12.25 & 5 & 64.0 & 51.0 & [b]\\
\hline
12. &HD 143006 & 2021/06/30 & VLT & IRDIS & K & 100 & 12.25 & 5 & 16.0 & 11.7 & [b] \\
13. &HD 143006 & 2021/07/22 & VLT & IRDIS & K & 100 & 12.25 & 5 & 16.0 & 34.0 & [b] \\
14. &HD 143006 & 2022/04/08 & Subaru & CHARIS & K & sat (81) & 16.20 & 4 & 60.5 & 79.7 & - \\
\hline
\end{tabular}
\tablecomments{Columns are target name, observation date, telescope used, instrument used, wavelength filter, inner working angle (radius), pixel scale, full width at half maximum of the stellar PSF, detector integration time, total integration time, and references. All FWHM measurement are rounded to the nearest pixel. For the non-coronagraphic observations IWA is determined through a visual identification of the PSF saturation boundary. In the case of HD 163296, this includes poorly resolved inner disk.} 
\tablenotetext{a}{\cite{Kusakabe2012}; 
                  \tablenotemark{b} \cite{Ren2023}; 
                  \tablenotemark{c} \cite{Garufi2024};
                  \tablenotemark{d} estimated from user manual}

\end{table*} 

\subsection{CHARIS}\label{subsec:charis}
Our CHARIS data were observed in high intensity non-coronagraphic mode at $K$-band wavelengths. We extracted the raw data cubes using the CHARIS Data Extraction Pipeline \citep[{\tt CHARIS-dep},][]{Brant2017}, choosing to over-sample. The CHARIS data cubes were then processed using the CHARIS Data Processing Pipeline \citep[{\tt CHARIS-dpp},][]{Currie2020}, with modifications to allow for non-coronagraphic data reduction. Since no sky frames were taken during the observation, sky subtraction was not performed. When performing cube registration, we added the {\tt method=1} tag to determine the star position using the centroiding method for an unmasked star. We also added the {\tt revise} flag for targets which struggled with finding an initial Gaussian fit for the peak brightness. We then performed spectrophotometric calibration using the Kurukz stellar library \citep{Kurucs1979} to fit the SED and passing the {\tt unsat} flag to allow for non-coronagraphic data processing. To match the HWP angles, we opted to use the angles registered in the HWP log files, rather than the HWP angles recorded in the {\tt fits} file headers, since HWP header recordings for observations taken before July 2022 are unreliable. 

In the final PDI calculations, we used the {\tt unres\_pol\_sub} flag to remove unresolved polarization which can contaminate the $Q_\phi$ and $U_\phi$ products. This unresolved polarization signal can originate from scattering by small dust grains around the central star, unresolved inner disk scattering, or instrumental polarization \citep{Avenhaus2014, Canovas2015}. We found this greatly improved image quality for non-coronagraphic CHARIS observations. During this stage we also used a Mueller Matrix to account for instrumental polarization
effects along the optical path \citep{vanHolstein2020,Joost2021}.

We successfully detect the disks of MWC 480, HD 163296 and HD 143006, but fail to detect any disk signatures around the six other targets. Observational information can be found in Table~\ref{tab:observations}. Figure \ref{fig:previously observed} shows our data reduction results for the three detected disks. Here we show the azimuthal Stokes vectors $Q_\phi$ and $U_\phi$ which represent signal and noise, respectively. Each disk shows a positive detection in $Q_\phi$ and expected noisy residuals in $U_\phi$. The $PI$, $Q$ and $U$ images can be found in Figure~\ref{fig:previously observed AP}. The disk structures will be discussed further in \S\ref{sec: analysis}. Our $Q_\phi$ and $U_\phi$ for the group of six new targets is shown in Figure~\ref{fig:CHARIS_PDI}. The $PI$, $Q$ and $U$ data can be found in Figure~\ref{fig:CHARIS_PDI AP}. No disk signatures are detected in these six observations. The point-source to the south in the HIP 80425 panel is a known stellar companion. The quadrupole patterns in $Q_\phi$ and $U_\phi$ are consistent with stellar polarization (likely originating from small dust grains around the central star) and an absence of polarization from an extended source. Theoretical approximations for PDI observations of a polarized point source can be found in Figure~\ref{fig:Theoretical Q U AP}, and will be discussed in \S\ref{sec: analysis}.

\subsection{fast-PDI} \label{subsec:fast-pdi}
Fast-PDI is a relatively new experimental addition to the SCExAO optical path \citep{Lozi2020}. It uses a dedicated IR camera (C-RED ONE) and a ferroelectric liquid crystal (FLC) modulator which rapidly switches polarization states in sync with the camera \citep{Lozi2020}. The combination of rapid FLC modulation, the Wollaston beam-splitter, and synchronized HWP cycles, allows a ``triple-differential imaging\footnote{\label{foot:Fast-PDi}\url{https://www.naoj.org/Projects/SCEXAO/scexaoWEB/030openuse.web/043fastPDI.web/indexm.html}}" polarimetric technique. This provides mitigation for instrumental polarization since any quasi-static polarization offsets can be quickly subtracted.

The fast-PDI data were taken simultaneously with the CHARIS data for each of our nine targets using ``Mode \#2" (see footnote~\ref{foot:Fast-PDi}) to allow for triple-differential imaging. As a result, all observations are non-coronagraphic except MWC 480, as with CHARIS. To reduce the raw data, we used the {\tt fastpdi\_dpp} data reduction pipeline. 
We first created the {\tt master dark} and {\tt master flat} files that are crucial for correcting saturated or dead pixels. We then used the pipeline to calibrate, register, and apply the PDI algorithm to extract the disk. 

We do not recover signal from the six CHARIS non-detections. In the three systems with CHARIS disk detections, the CHARIS data achieve higher SNR than the fast-PDI data. Therefore, we do not pursue the fast-PDI results further in the scientific analysis. For completeness, we show one example of a fast-PDI result in Figure~\ref{fig:fast_PDI}. The eastern side of the NIR inner disk is clearly detected, with the peak signal in the SE, as with CHARIS. However, the outer ring that is visible with CHARIS is not resolved with fast-PDI (Figure~\ref{fig:SNR HD143006 AP}). 

\section{Analysis of the Detected Disks}\label{sec: analysis} In this section we analyze our observational results and compare to previous PDI imaging where available. Section \ref{subsec: SNR} describes how we obtained signal-to-noise ratio (SNR) images and discusses the significance of our detections. Sections \ref{subsec: mwc480}, \ref{subsec:HD163296 analysis} and \ref{subsec: HD143006} analyze any temporal variation seen in MWC 480, HD163296 and HD143006, respectively, by comparing to PDI imaging done with VLT/SPHERE/IRDIS. Finally, in section \ref{subsec: non detections pol signal} we analyze the six targets where disks were not successfully detected and describe what can be understood from their polarization signal. 

\subsection{SNR Assessment} \label{subsec: SNR}
The SNR for each of the three previously detected disks is shown in the right column of Figure \ref{fig:previously observed}. Noise was calculated following \cite{Dykes2025} by separating the $U_\phi$ image into annuli of FWHM width (see Table~\ref{tab:obs info} for FWHM in pixels), and calculating the robust standard deviation (STD, $\sigma$) within each annulus. Robust STD uses median absolute deviation (MAD$=\rm{median}(|X_i - \tilde X_i|)$) and is given by
\begin{equation} \label{eq:mad std}
    \sigma \approx \frac{\rm{median}(|X_i - \tilde X_i|)}{\Phi^{-1}(3/4)} 
    \approx 1.48 \times\rm MAD
\end{equation}
where $\tilde X_i$ is the median of the data and $\Phi^{-1}(P)$ is the normal inverse cumulative distribution function evaluated at probability $P$. This method was chosen over traditional STD due to its robustness against outliers.  

Once the noise at one radial separation is calculated, a new annulus of the same width is created shifted 1 pixel radially outwards, where robust STD is again calculated until the edge of the field of view (FOV) is reached. These $U_\phi$ noise annuli are combined to from a new ``noise image". SNR maps were obtained by dividing the 
$Q_\phi$ images by the noise images obtained using Equation~\ref{eq:mad std}. We most strongly detect the inner disk of HD 143006 with a peak SNR of 15 while the other disks are detected with an overall local SNR of $\sim$3-10. 

\subsection{MWC 480} \label{subsec: mwc480}
\begin{figure*} [!htpb]
    \centering
    \includegraphics[width=1\textwidth]{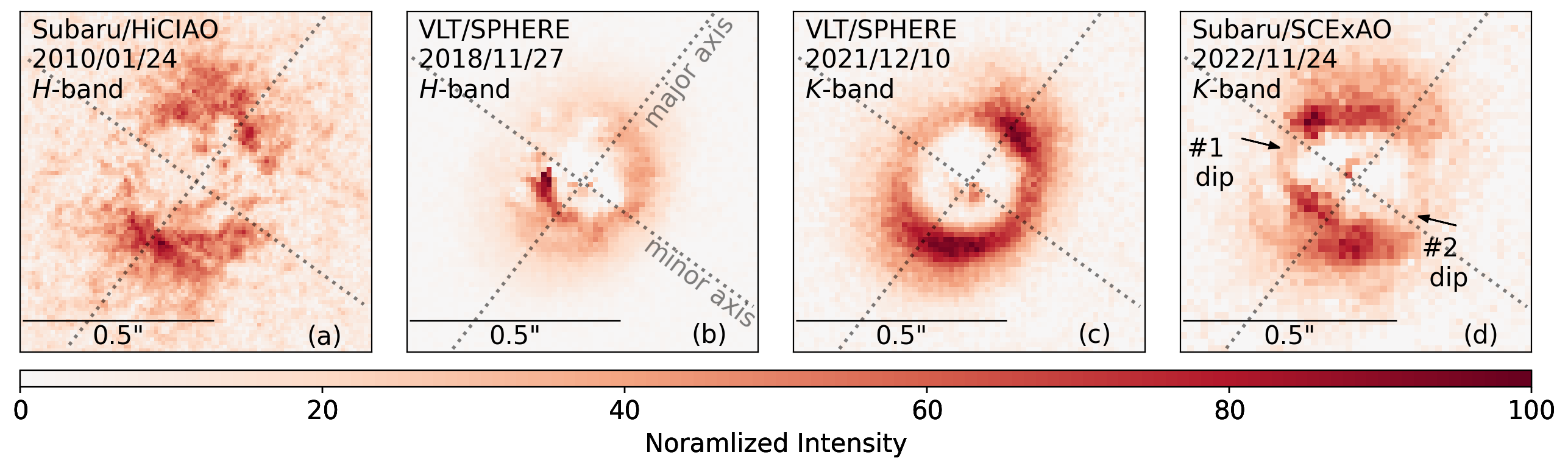}
    \caption{Four epochs of MWC 480 observations: 
    \cite{Kusakabe2012} HiCIAO 2010/01/24; \cite{Garufi2024} SPHERE 2018/11/27; \cite{Ren2023} SPHERE 2021/12/10; this work SCExAO/CHARIS 2022/1/24. 
    All images are north-up, east-left. The images are normalized to the peak brightness. The original sign of the pixel values is conserved. The HiCIAO observation has no available $Q_\phi$ data so the $PI$ product is shown. We note two shadow-like brightness dips in our SCExAO observation. Major (\ang{143}) and minor (\ang{53}) axes  -- determined by the disk model fitting (\ref{subsec: mwc480}) -- are indicated by the dotted lines. Note, all angles are measured here with north-up as \ang{0}. 
    }
    \label{fig: MWC 480 4}
\end{figure*} 

\begin{figure*} [!htpb]
    \centering
    \includegraphics[width=0.75\textwidth]{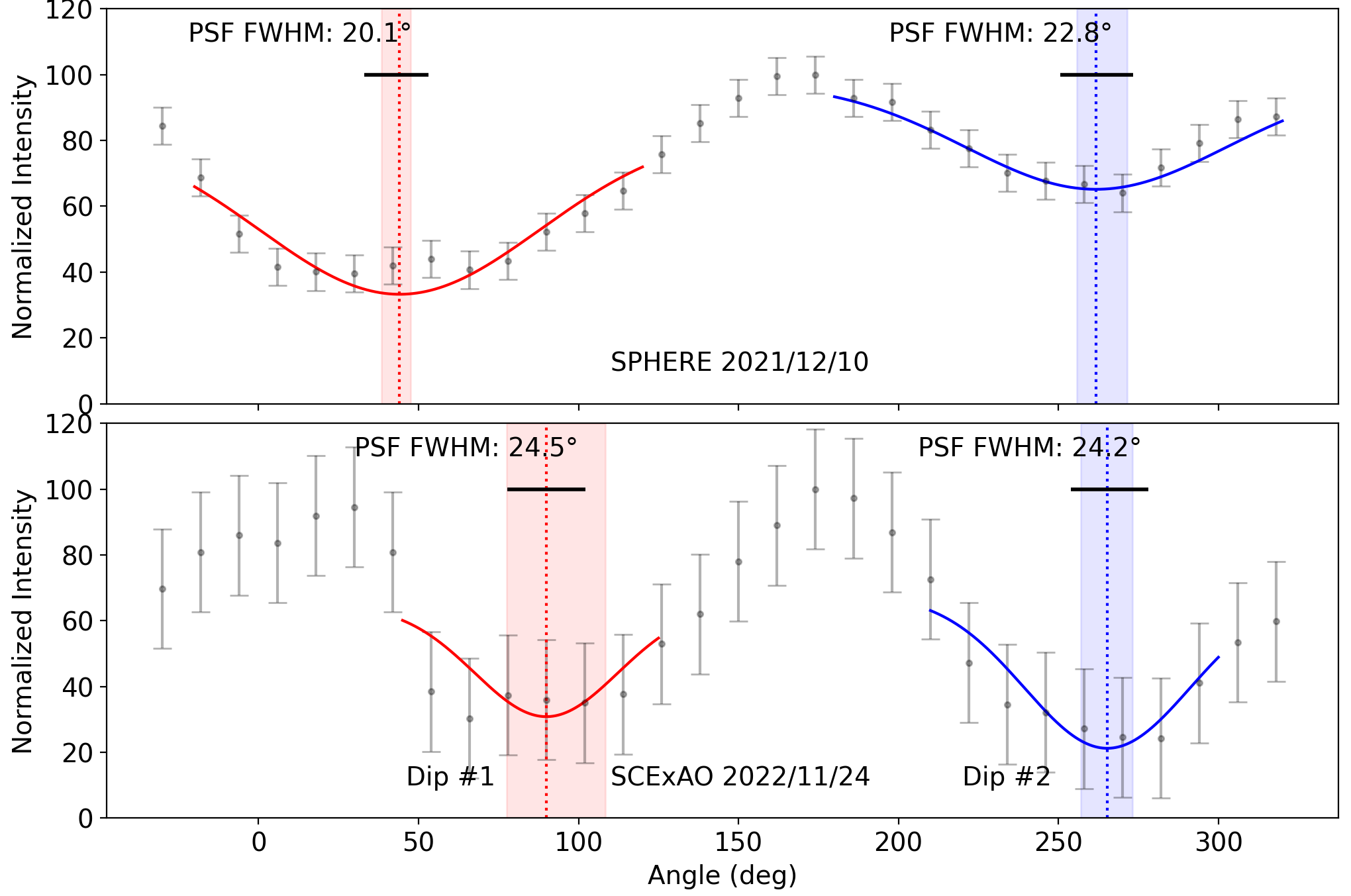}
    \caption{Azimuthal brightness profiles measured at a deprojected radius of 0\farcs19 for two epochs of MWC 480 PDI observations. The y-axis is in the same normalized intensity units as Figure~\ref{fig: MWC 480 4}. The red curves are a gaussian fit (equation \ref{eq: gauss}) to the brightness dips on the east side of the image (Dip \#1), while the blue curves are the gaussian fit to the west (Dip \#2). The PSF FWHM bars represent the angular resolution at the position of the gaussian fit centers. The dotted lines show the centers of each brightness dip with the shading showing the $1\sigma$ errors. 
    }
    \label{fig: MWC profile}
\end{figure*} 

In this section, we compare our results for MWC 480 to three previously published IR observations taken with similar resolution to our own and also using PDI. A summary of PDI disk observations can be found in the first block in Table~\ref{tab:obs info}. A gallery of the disks is provided in Figure~\ref{fig: MWC 480 4}. The first NIR PDI observations of this target were taken on 2010 January 24 using the Subaru high-contrast imaging instrument (HiCIAO) in $H$-band \citep{Kusakabe2012}. The authors noted a dark lane along the minor axis of ~\ang{57} w.r.t North, which they attributed to depolarization due to the grain scattering phase function, and not an indication of material depletion. The full ring structures have been revealed in mm continuum, showing no material depletion along this axis \citep{Long2018}. We note, that we find the minor axis to be PA=\ang{53} in our SCExAO data (see Figure~\ref{fig: ring fit MWC 480}), or \ang{4} offset from \cite{Kusakabe2012}, likely due to our access to better resolution of the disk geometry. The disk was observed again on  2018 November 27 using VLT/SPHERE/IRDIS in $H$-band, with the observations later published by \cite{Garufi2024}. The authors note the global self-shadowing of the disk. Another set of observations on 2021 December 10 with the 185 mas Lyot stop were published by \cite{Ren2023} in $K$-band using VLT/SPHERE/IRDIS.

\begin{table} [!htbp]
\centering
\caption{MWC 480 Brightness Dip Fitting Results }
\label{tab:mwc480 params}
\begin{tabular}{l@{\hspace{1mm}}c@{\hspace{1mm}}c@{\hspace{1mm}}c@{\hspace{2mm}}c@{\hspace{1mm}}c}
\toprule
 & Dip \#1 & Dip \#2 & Dip \#1 & Dip \#2 \\
& 2022/11/24 & 2022/11/24 & 2021/12/10 & 2021/12/10 
\\
\midrule
     A [\%] & $-33.8^{+14.8}_{-13.7}$ & $-46.7^{+11.9}_{-12.2}$ & $-49.8^{+4.1}_{-32.1}$ & $-32.5^{+4.7}_{-3.4}$ 
     \\
   $\mu$ [deg] &  $89.9^{+18.4}_{-12.4}$ &   $265.3^{+7.9}_{-8.4}$ &   $44.1^{+3.4}_{-5.7}$ & $261.8^{+9.6}_{-5.9}$ 
   \\
$\delta$ [deg] &    $44.8^{+16.6}_{-14.8}$ &    $51.8^{+12.4}_{-10.8}$ &   $87.6^{+12.4}_{-7.6}$ &  $81.4^{+11.0}_{-17.6}$  
\\
     C [\%] &    $64.6^{+4.1}_{-4.0}$ &    $67.9^{+4.4}_{-4.1}$ &  $83.1^{+15.8}_{-2.0}$ & $97.6^{+16.9}_{-3.4}$
     \\
\bottomrule
\end{tabular}
\tablecomments{
All parameters and error bars were derived using the {\tt emcee} Python package to fit gaussian curves (equation \ref{eq: gauss}) to Dips \#1 and \#2.}
\end{table}

For the purposes of our analysis, we only consider the $K$-band SPHERE epoch to not introduce wavelength dependencies. The $J$, $H$ and broadband $JHK$ $Q_\phi$ images can be found in Figure~\ref{fig: color_compare_MWC480}. 
For the remainder of our analysis, we focus on the two shadow-like brightness dips along the minor axis of our SCExAO observation, which are indicated in Figure~\ref{fig: MWC 480 4}. A similar pattern of brightness decrease is seen in the HiCIAO observation, and brightness decrease along the backside (NE) of both SPHERE epochs. 

To quantify the azimuthal light distribution in the disk, we begin by fitting a ring model to the SPHERE 2021 $Q_\phi$ observation (in a similar manner to \citealt{Benisty2017}).  The SPHERE 2021 epoch was selected to fit the disk model since it is the epoch with the best defined elliptical disk shape. The model is a simple 2D gaussian ring with a defined position angle (PA), inclination, width, radial separation, and center. The model ring is fit using Markov chain Monte Carlo (MCMC) with the python {\tt emcee} package \citep{Foreman2013}.  The fitted ring assumes a circular disk projected at an inclination. We used 6 free parameters in {\tt emcee}: radial separation (R0), ring width, inclination (inc), PA, and central star position (x0, y0). We used 60 walkers, and 4000 iterations resulting in an acceptance fraction of 0.5. The noise is defined as the standard deviation in the $U_\phi$ data. The resulting disk region can be found in Figure~\ref{fig: ring fit MWC 480}. We determine the PA to be \ang{142}.$6^{+0.5}_{-0.5}$, inclination to be \ang{36}$.9^{+0.3}_{-0.3}$ and R0 to be 0\farcs$19$. 

To quantify the location of the brightness dips and potential shadow movement, we fit azimuthal brightness profiles to the three epochs in question. We fit the three azimuthal profiles along an elliptical path in the $Q_\phi$ data defined by the inc, PA, and median R0 obtained from the fitting above. We then sample the \ang{360} ellipse using 60 FWHM  apertures. The choice of 60 apertures follows Nyquist-Shannon sampling theorem \citep{Shannon1949} such that each aperture has 1/2 overlap with its neighbor. The brightness value represented by each point on the azimuthal profile is taken to be the mean value within the corresponding FWHM aperture. Following a similar method to \cite{Benisty2017} and \cite{Debes2017}, we then follow the same azimuthal path in the $U_\phi$ data and take the standard deviation of the resulting $U_\phi$ azimuthal profile to represent the noise in the azimuthal fit.

To quantify the azimuthal location of the brightness dips at each epoch, we fit a gaussian function, 
\begin{equation} \label{eq: gauss}
    f(\theta) = A\times \exp{\left(-\left(\frac{(\theta-\mu)}{2\delta}\right)^2\right)}+C
\end{equation} 
to the two dips in each epoch using MCMC, where $\theta$ represents an azimuthal angle. We used four free parameters: amplitude (A), center ($\mu$), width ($\delta$), and offset (C), 40 walkers and 4000 steps for each fit. We supplied the noise calculated with the $U_\phi$ images to the likelihood function. The resulting fitted gaussian dips can be seen in Figure~\ref{fig: MWC profile}, and the best fit parameters are provided in Table~\ref{tab:mwc480 params}. The azimuthal widths of the dips in the fitting results exceed the PSF FWHM, indicating that the dips are azimuthally resolved.

As is apparent in Figure~\ref{fig: MWC profile}, the dips move by different degrees. Dip \#1 showed a $3.6\sigma$ counter-clock-wise (CCW) change in apparent azimuthal center location corresponding to a movement of 
46$\pm$\ang{13} in 0.96 years.
The dip also narrowed in the azimuthal width between 2021 and 2022, as shown by the $2.3\sigma$ shrinkage in the width parameter. Following the same trend, Dip \#2 shows a slight CCW movement in center, however the two values are within $1\sigma$ of one another, and thus not statistically significant. We shall discuss the potential causes of these movements in \S\ref{sec:Discussion}.

\begin{figure}[!htpb]
    \centering
    \includegraphics[width=0.44\textwidth]{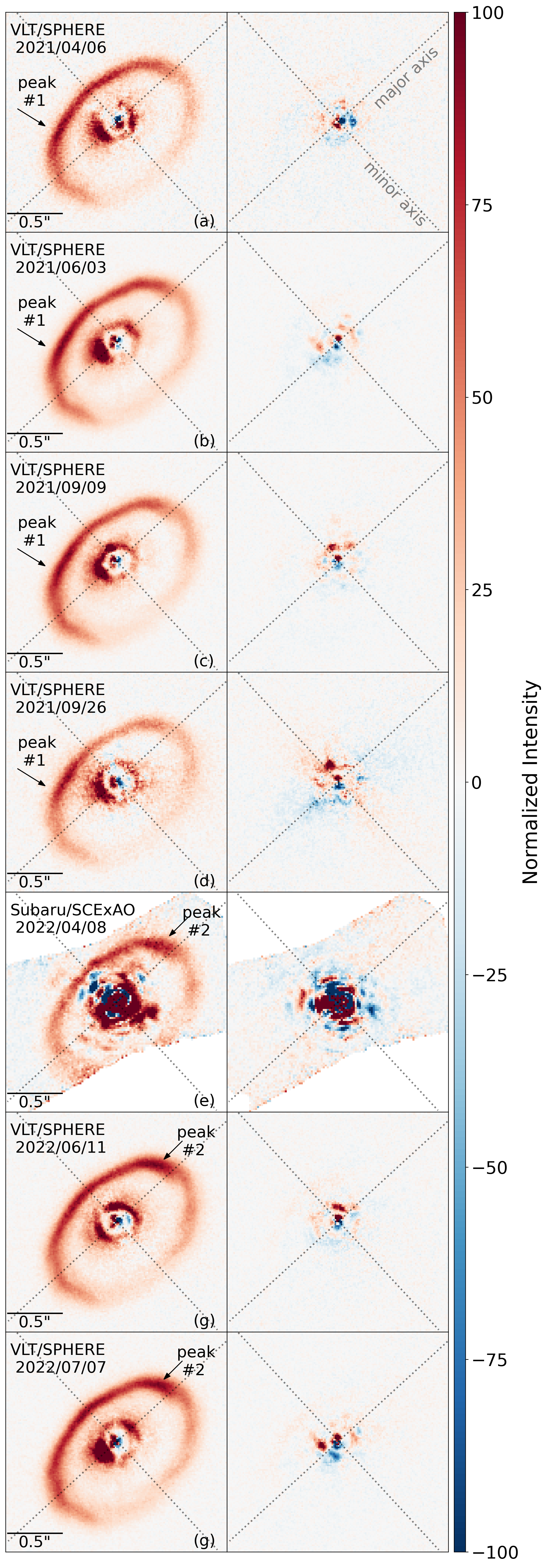}
    \caption{SPHERE \citep{Ren2023} and SCExAO $K$-band $Q_\phi$ (left) and $U_\phi$ (right) for HD 163296. Panels are normalized to the brightest point in the outer ring.} 
    \label{fig:HD163296 SPHERE}
\end{figure}

\begin{figure*}[!htpb]
    \centering
    \includegraphics[width=0.99\textwidth]{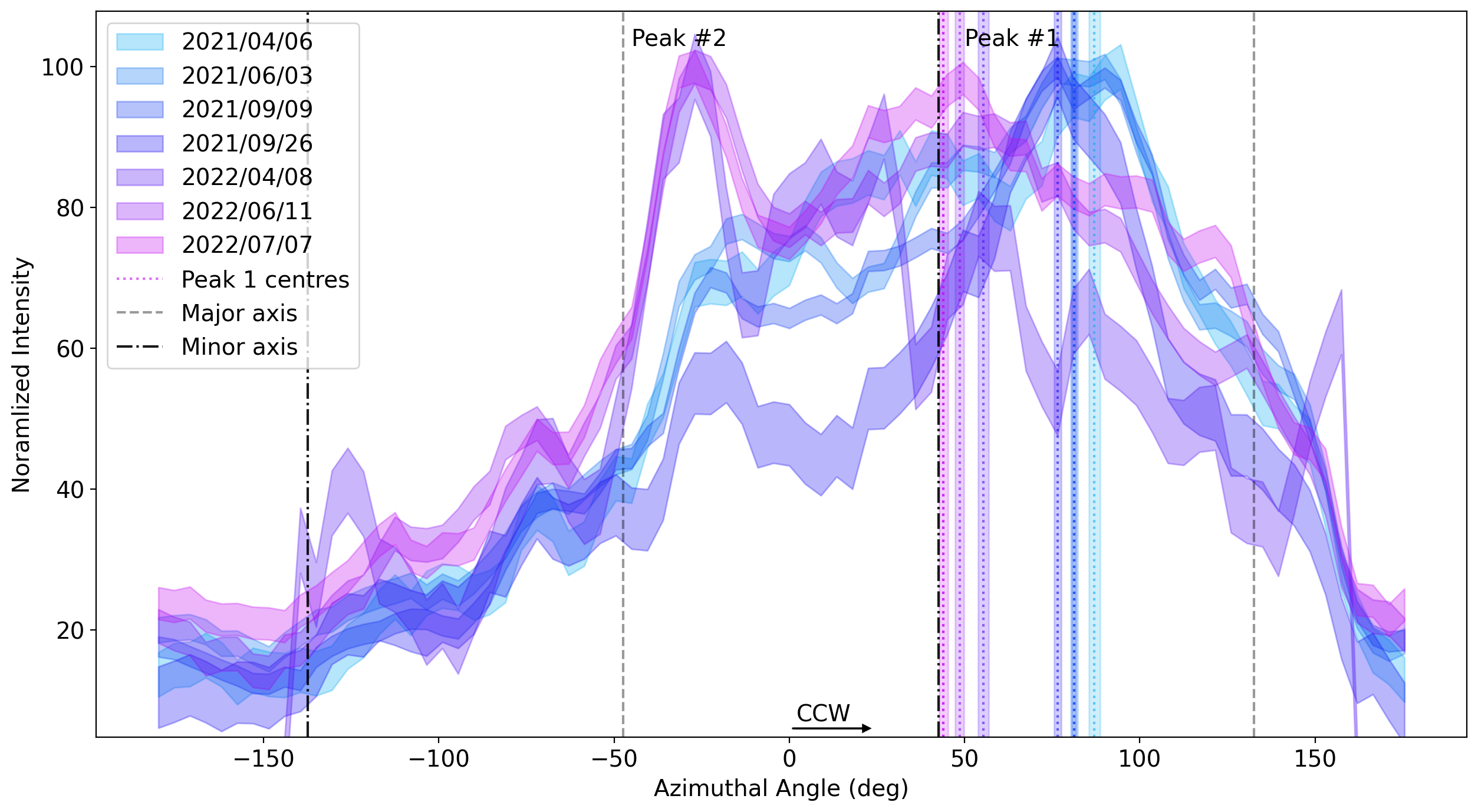}
    \caption{ Azimuthal brightness profiles measured at a deprojected radius of 0\farcs64 for the 7 $K$-band epochs of HD 163296. Each profile is normalized. The location of the peak in each profile is traced by the vertical dotted lines. We note the gradual shift of the SE peak (Peak \#1) and the consistency of the NW peak (Peak \#2). Due to CHARIS' limited FOV, the data SCExAO profile between \ang{162} and \ang{-144} is set to 0. } 
    \label{fig:HD163296 All}
\end{figure*}

\begin{figure}[!htpb]%
    \centering
    \includegraphics[width=0.45\textwidth]{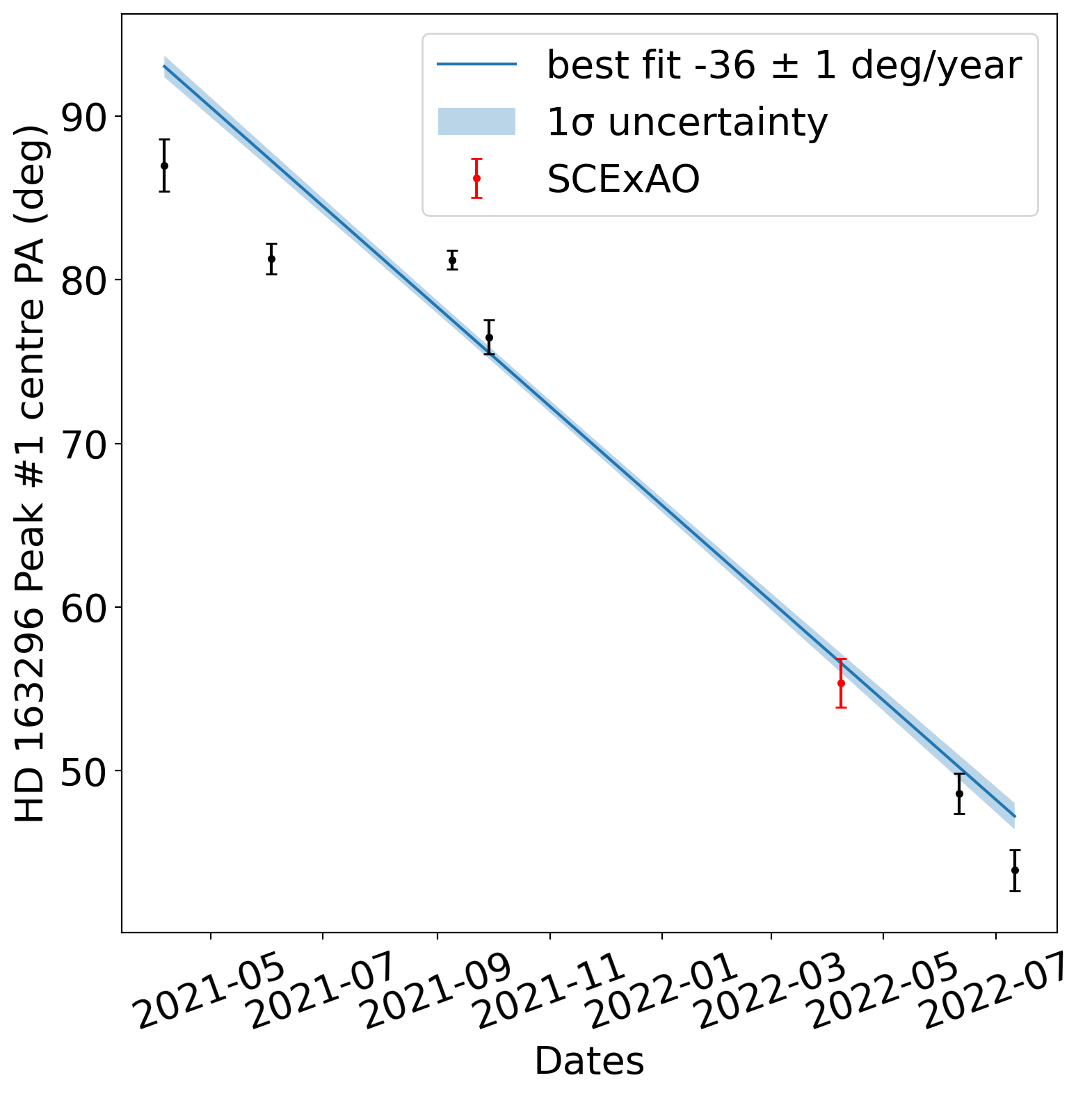}
    \caption{Azimuthal brightness  movement of HD 163296 Peak \#1 over time. The peak brightness locations were determined by fitting the peak with MCMC. The linear-best fit line shows an average rate of change in peak location which corresponds to a $R^2=0.95$ coefficient.} 
    \label{fig:HD163296 Move}
\end{figure}

\subsection{HD 163296} \label{subsec:HD163296 analysis}
HD 163296 has been observed in $K$-band over six epochs with SPHERE/IRDIS using the 185 mas Lyot stop, spanning from 2021 June 4 to 2022 July 7. The ring at $\sim$60au is detected with SCExAO as with SPHERE. The coronagraphic SPHERE images also reveal the $\sim$15\,au inner ring, however this region is overwhelmed by stellar PSF in the SCExAO observations and thus not part of our analysis. 

The $Q_\phi$ and $U_\phi$ panels for each SPHERE epoch can be seen in Figure~\ref{fig:HD163296 SPHERE} as well as our SCExAO epoch. It is apparent that the azimuthal brightness varies between epochs. The global brightness peak shifts from SE (Peak \#1) to NW (Peak \#2) in the time between the 2021/09/26 and 2022/04/08 epochs and remains NW in subsequent epochs. To determine whether the shift is gradual or abrupt, we extract azimuthal profiles along the visible outer ring in each epoch, which is determined using the same method described in section \ref{subsec: mwc480} where MCMC is used to determine the best ring fit (see Figure~\ref{fig: ring fit MWC 480}). The resulting profiles are shown in Figure~\ref{fig:HD163296 All}. As expected from the disk images, two peaks are apparent at PA$\sim$\ang{90} (Peak \#1) and PA$\sim$\ang{-27} (Peak \#2), with the Peak \#2 in the NW appearing only after the 2021/09/26 epoch. 

However, while Peak \#2 appears relatively stationary, Peak \#1 shows a potential migration CW. To test the significance of this migration, we fit a gaussian to the eastern peak in each epoch's azimuthal profile (Figure~\ref{fig:HD163296 All}) to determine its peak location using the same MCMC method described for the MWC 480 dip fitting. The PA of Peak \#1's versus time is shown in Figure~\ref{fig:HD163296 Move}. 
Using {\tt SciPy curve\_fit} we fit a linear model to the points to assess if the rate of change is consistent, resulting in a line of best fit with a slope of \ang{-36}/year $\pm$\ang{1}/year. This corresponds to an $R^2$ value of 0.95, indicating high correlation between the data and line of best fit.
The fit suggests that the peak moves at a linear rate, with the caveat that the mean values do not incorporate the disk scattering background, which will be discussed more in section \ref{subsec:background}. 

\begin{figure*}[!htpb]
    \centering
    \includegraphics[width=0.95\textwidth]
    {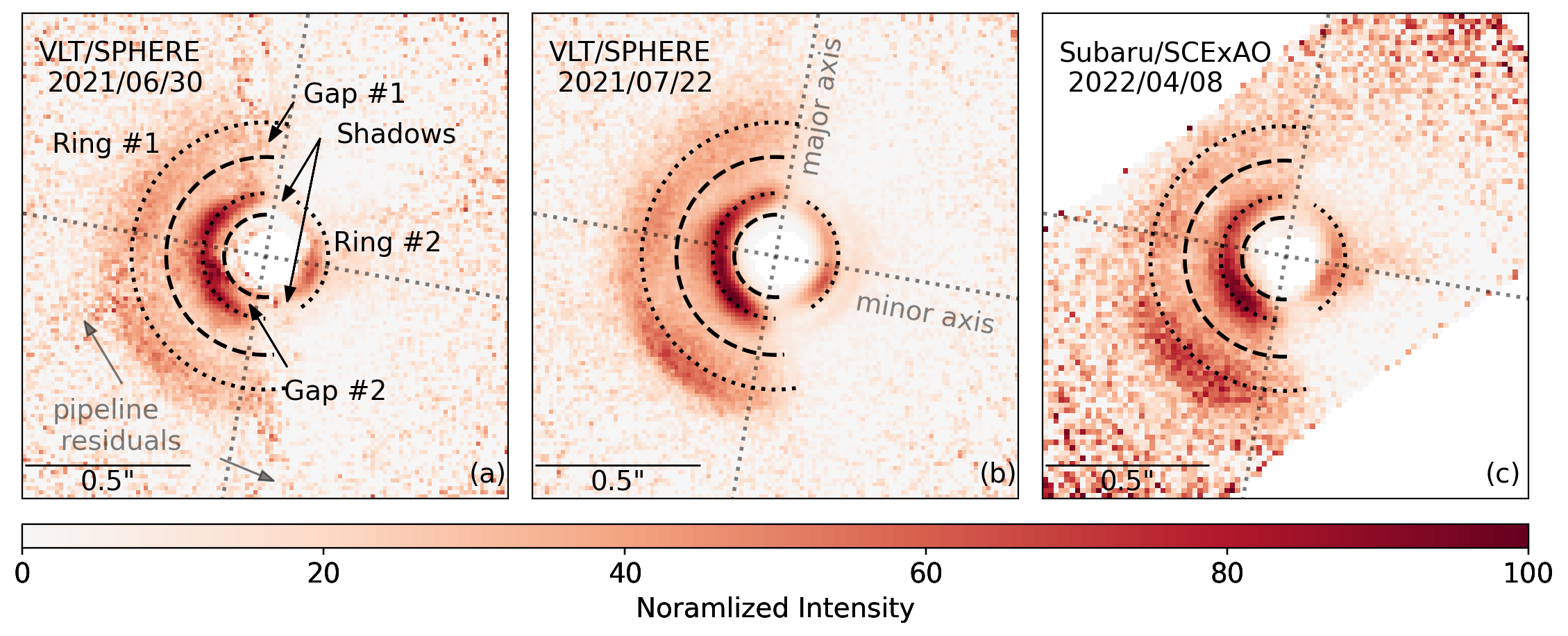}
    \caption{HD 143006 $K$-band PDI epochs showing VLT/SPHERE/IRDIS (panels (a) and (b)) and Subaru/SCExAO/CHARIS (c). All images are scaled as $r^2$ (with $r$ defined in the sky plane due to the disk's minimal inclination) to enhance faint the faint outer ring. We use the same annotation as \cite{Benisty2018} to denote the visible rings and gaps as well as shadowing. The major \citep[\ang{170},][]{Benisty2018} and minor (\ang{80}) axes are denoted by the dashed lines.
    } 
    \label{fig: HD143006 k band}
\end{figure*} 

\begin{figure*}[!htpb]
    \centering
    \includegraphics[width=0.9\textwidth]{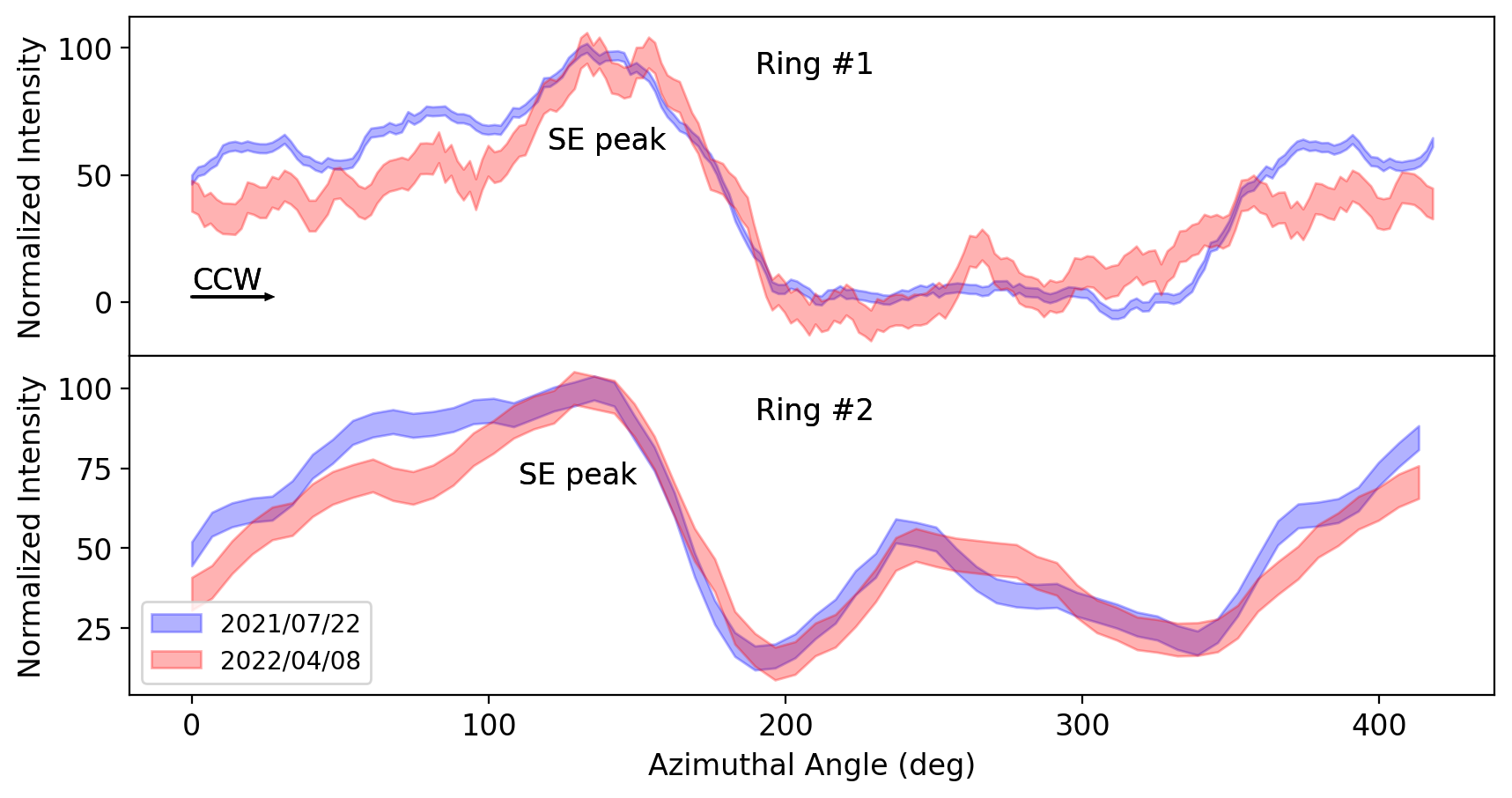}
    \caption{HD 143006 azimuthal profiles for the Ring \#2 (top; measured at a deprojected radius of 0\farcs18) and Ring \#1 (bottom; measured at a deprojected radius of 0\farcs4). Azimuthal angle is measured with respect to \ang{0} north. The shading represents the error bars on each profile.
    }
    \label{fig: HD143006 azimuthal}                          
\end{figure*} 

Additionally, we investigated whether azimuthal brightness variations correspond to changes in the ring’s radial width, but found no statistically significant width variation between epochs.

\subsection{HD 143006} \label{subsec: HD143006}
HD 143006 has past PDI observations with VLT/SPHERE/IRDIS in $J$-band \citep{Benisty2018} and $K$-band with the 185 mas Lyot stop \citep{Ren2023}. Figure~\ref{fig: HD143006 k band} shows our SCExAO $Q_\phi$ data and the SPHERE $K$-band epochs, $r^2$-scaled to match Figure 2 from \cite{Benisty2018}, with the corresponding disk features outlined. The 2021 June SPHERE observation contains a crosshair pattern from pipeline residuals and thus will not be included in the disk structure analysis. 

In all epochs, we detect the same key disk features as \cite{Benisty2018}, such as Ring \#1, Ring \#2, Gap \#1, Gap \#2, and the two shadow lanes bisecting Ring \#2 at a PA of \ang{190} and \ang{340}. Similarly to \cite{Benisty2018}, we detect the brightest signal in the SE quadrant on Ring \#2.  \cite{Benisty2018} suggested that an inclined inner disk is responsible for the obscuring of the western side of the outer disk and the shadow lanes. 

To assess for any azimuthal variation we took azimuthal profiles of both rings following the same methods used for MWC 480 and HD 163296. The profiles are shown in Figure~\ref{fig: HD143006 azimuthal} for the 2021 July SPHERE and 2022 SCExAO epochs. The SE peak aligns for both epochs in both rings, as do the visible outer disk boundaries and shadow lanes. Peak \#1 coincides with a bright clump seen in mm continuum \citep{Perez2018}, suggesting that this IR brightness peak is correlated to a concentration of mm-sized dust. Notably, the kinematically predicted planet in \cite{Pinte2020} resides close to the SE peak in Ring \#2, as shown in Figure~\ref{fig:previously observed}.

In Ring \#2, the positions of the shadows lanes appear constant between the SPHERE and SCExAO epochs, also suggesting no significant change in the inner disk over the 9 month period. A fit to both shadows in ring 2 using MCMC showed $1\sigma$ agreement between the two epochs. 

Similarly to HD 163296, we investigated the radial widths of both rings  and Gap \#1 , and found that their widths remain consistent over time. The width and position of Gap \#1 are relatively constant with $1\sigma$ consistency across epochs and azimuthal angles. 

\subsection{Polarization Signal in Non-detections}
\label{subsec: non detections pol signal}
\begin{table}
\centering
\caption{Polarization Properties of Targets Without Disk Detections}
\label{tab: pol frac}
\begin{tabular}
{lcc}
\toprule
Target & AoLP [deg] & Polarization Fraction [\%] \\
\midrule
HD 56895  & 45 & 0.7     
\\
HD 144432 & 53 & 1.5
\\
HIP 80425 & 79 & 0.5
\\
HIP 81474 & 131 & 0.8
\\
HD 148352 & 99 & 0.6
\\
PDS 76 & 159 & 1.3
\\
\bottomrule
\end{tabular}
\tablecomments{Wavelength averaged angle of linear polarization (AoLP) and polarization fraction in percentage for all 6 non-detections. Reference models are found in Figure~\ref{fig:Theoretical Q U AP}. 
}
\end{table}
\begin{figure*}[!htpb]%
    \centering
    \includegraphics[width=0.99\textwidth]{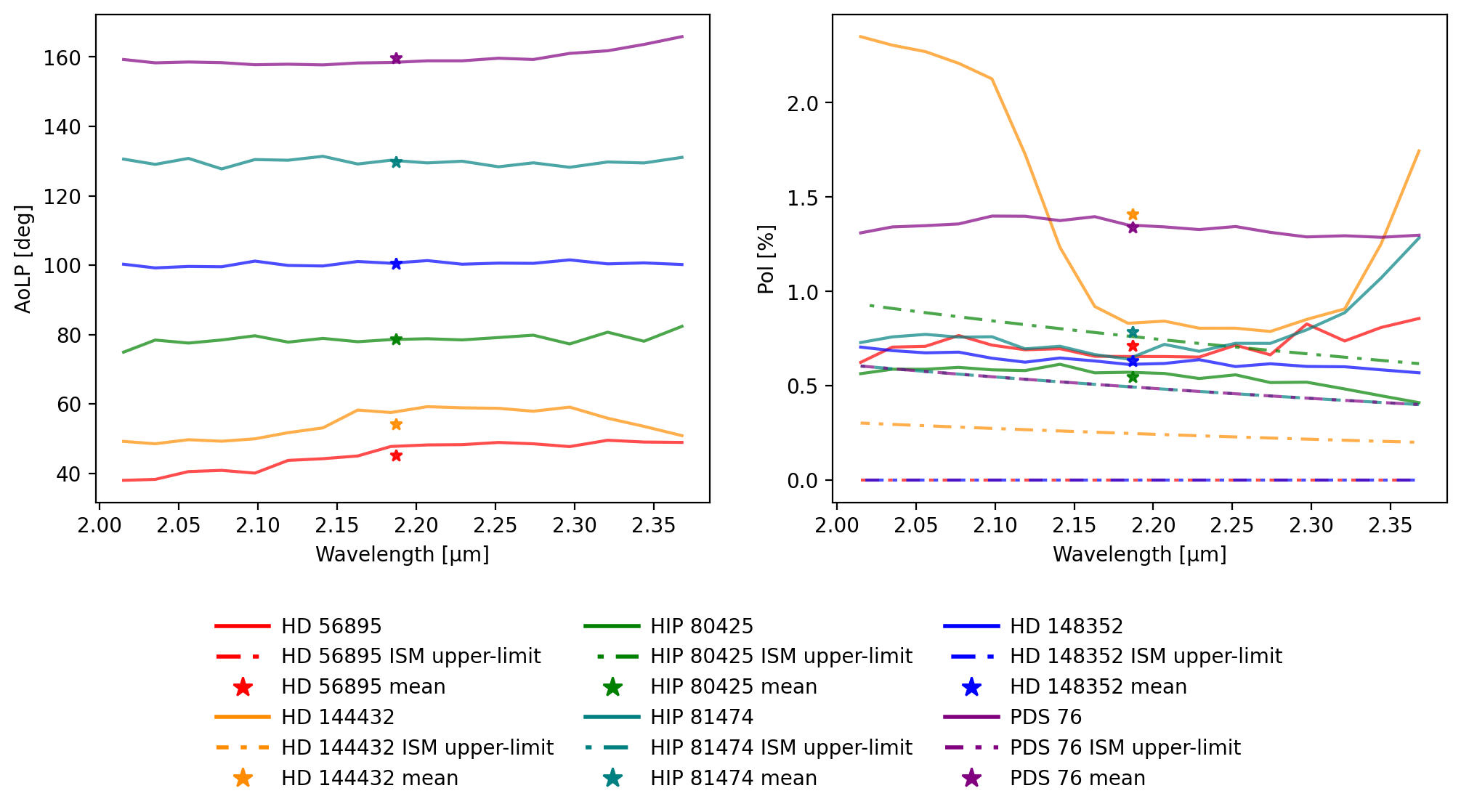}
    \caption{ Polarization angle (left) and fraction (right) as a function of wavelength for the 6 non-detections. The star shapes indicate the mean values reported in Table~\ref{tab: pol frac}. The dash-dot lines in the right panel indicate the theoretical maximum polarization percentage if polarization is purely ISM caused, as per equation \ref{eq:serkowski}. HD~56895 and HD~148352 have reported extinctions of 0, and thus would have zero ISM polarization contribution.}
    \label{fig:AOLP and percent}
\end{figure*}

Each of our observations of the remaining six targets are consistent with a slightly polarized star and an absence of detectable extended structure. By applying a FWHM aperture to the calibrated intermediate products $Q^+_R$, $Q^+_L$, etc., performing aperture photometry, and then applying double differentiation (equations \ref{eq:stokes first} to \ref{eq:uphi}), we can determine the wavelength averaged angle of linear polarization (AoLP) and polarization fraction for each target. The results of these calculations are found in Table~\ref{tab: pol frac}. Most targets are $<1\%$ polarized, except PDS 76 and HD 144432 which are $1.3\%$ and 1.5\% polarized respectively.

To demonstrate that our results are consistent with polarized point sources in the absence of extended structure, we use analytical models to showcase that the quadrupole patterns seen in $Q_\phi$ and $U_\phi$ naturally arise when a polarized star undergoes double differential imaging. These models are shown in Figure~\ref{fig:Theoretical Q U AP}. Each row shows the analytical double-differentiation result for synthetic $Q^\pm_{R/L}$ and $U^\pm_{R/L}$ noisy PSFs with the respective target's measured polarization fraction applied in the appropriate direction. As an example, for HD~144432 we measured polarization fractions of $0.4$\% in $Q^-_R$ and 1.4\% polarization in $U^+_R$. Applying these values to the model PSFs and performing double-differentiation produces the results shown in the second row of Figure~\ref{fig:Theoretical Q U AP}. The resulting sign and degree of intensity of the $Q$ and $U$ signals, as well as the orientation of the quadrupole patterns in $Q_\phi$ and $U_\phi$ match the CHARIS products shown in Figure~\ref{fig:CHARIS_PDI AP}.

While these patterns are consistent with a non-detection of extended emission, they do not provide any indication of why the surrounding disks could not be detected. Self-shadowing or strong dust settling could be responsible for some of these non-detections, as well as Meeus group being a potential indicator for detachability which we will discuss further in section \ref{subsec:non-detections discussion}.

To investigate the potential polarization origin in each target, we plot the AoLP and polarization percentage as a function of wavelength in Figure~\ref{fig:AOLP and percent}. In the right panel, we compare our polarization percentage for each target with the theoretical upper limit in ISM dominated polarization. ISM polarization is correlated with extinction magnitude \citep{Serkowski1975}. Five of our non-detections have reported finite extinction magnitudes, while HD~148352 and HD~56895 have a reported extinction of 0 \citep{Chen2012,Guzman2021}.

For the targets with finite extinction, a theoretical maximum polarization percentage can be obtained from the relation in \cite{Serkowski1975}, $P_{max}(\%) = 9.0E_{B-V}$ , which is then used in combination with the Serkowski law, 
\begin{equation} \label{eq:serkowski}
    P(\lambda) = P_{max}\exp\left[-K\ln^2{\left(\frac{\lambda_{max}}{\lambda}\right)}\right]
\end{equation}
to obtain wavelength dependent maximum polarization percentage. Here, $K=0.01+1.66\lambda_{max}$ \citep{Wittet1992}, and $\lambda_{max}$ is assumed to be 0.55$\mu$m.

As shown in Figure~\ref{fig:AOLP and percent}, HP 80425 exhibits a polarization fraction lower than the theoretical ISM maxima, suggesting that the observed polarization may be dominated by interstellar polarization. For the other sources, the polarization exceeds the theoretical ISM maximum, and thus other contributors such as circumstellar material must contribute to the observed polarization.

HD~144432 shows a statistically significant variation in polarization with respect to wavelength, with an amplitude of $\sim$1.6\%, compared to typical variations of 0.1\%-0.2\% within our sample. The corresponding AoLP shows a variation of \ang{8},
which lies within one standard deviation of the sample mean AoLP variation. This large chromaticity in polarization fraction could be astrophysical if the relative contributions of polarized scattered light and weakly (or un-)polarized emission from hot dust and/or inner-disk gas vary across the bandpass.

\section{Discussion} \label{sec:Discussion}
Building on the observational results presented in \S\ref{sec: analysis}, we now interpret the variability signatures and non-detections in the context of disk structure, 
and sample properties. In section \ref{subsec:movement} we discuss the temporal variability and their possible causes. In section \ref{subsec:background} we explain our constraints in understanding the background scattering. 
Finally, in section \ref{subsec:non-detections discussion} we discuss the implications of our non-detections and any possible trends that could indicate why six of our nine target disks were not detected. 

\subsection{Movement of Brightness Dips}\label{subsec:movement} 
We observe a diverse range of shadow and light variability across our sample of three detected disks. MWC 480 has a pair of roughly symmetric brightness dips which could potentially be shadows cast by an inner disk at least modestly misaligned with the outer disk, and one of the two shows $3.6\sigma$ movement between epochs. HD 1634296 does not show any potential shadow lanes, but instead an apparently linear progression of the brightness peak. In contrast, HD 143006 shows no detectable change in shadow pattern between epochs, suggesting that if the pair of symmetric shadows are caused by a highly misaligned inner disk, it is relatively stationary over the timescales we probe. There are potential minor fluctuations in local scattering patterns in the NE, however the SE brightness peaks and shadow lanes remain static.

To discuss the origin of these changes in illumination, we first determine the rate of movement. The Keplerian orbital frequency for an object on a circular orbit within a protoplanetary disk is 
\begin{equation} \label{eq: speed}
    \omega = \sqrt{\frac{M_\star/M_\odot}{(r/{\rm AU})^3}}\times360 \degree/{\rm year}
\end{equation} 
where $M_\star$ is the stellar mass and $r$ is the radial separation.

Applying MWC 480's stellar mass  ($2.1M_\odot$), and the radial separation of the visible ring (30au) to equation \ref{eq: speed}, results in a velocity of \ang{3.06}/yr. This is significantly slower than the $48\pm$\ang{13}/year movement of Dip \#1 on the ring between the 2022 SCExAO and 2021 SPHERE observations, suggesting that this is not a movement of material within the visible ring, but an illumination pattern change. 

Similarly, we can ascertain that the changes in brightness in the HD 163296 disk are not related to material moving within the visible ring. The average rate of change between epochs is 36$\pm$\ang{1}/year, far exceeding the local Keplerian velocity of \ang{1.08}/year. This rate of change is consistent with Keplerian movement of inner disk material at $\sim$6 AU. 
However, an orbiting inner object is generally associated with a moving shadow cast on the outer disk, rather than a moving patch of brightness. Changes in brightness without discernible shadowing could indicate local changes in scale height and/or scattering. However, the linear rate of change and propagation in a singular direction would suggest these changes are not random. It is notable that one of the \cite{Pinte2020} kinematically predicted planets (the inner one) is located near Peak \#1. 
However, Peak \#1's super-Keplerian rate of motion argues against any physical association with the proposed planet.

We cannot provide a robust speculation of the origin of these illumination changes in MWC 480 and HD 163296 without detailed disk modeling. The morphology of the brightness dips in MWC 480 could suggest precession of one or more misaligned inner disks, though we cannot determine in what plane the disk is processing such that the dips do not move together (e.g. \citealt{Debes2017}, \citealt{Pinilla2018}, \citealt{Stolker2017}). \cite{debes2023} found two shadows moving at different rates in the TW Hydrae disk, and concluded that this was most likely caused by two inclined inner disks with different precession timescales. MWC 480 may also possess two or more inner disks with different precession timescales. We cannot at present confirm the existence of multiple inner disks, and detailed modeling is out of scope of this paper. 

We also caution that other factors can create a de-illumination pattern which mimics shadows. Depolarization is expected along the disk minor-axis \citep{Muto2012,Hunziker2021, Tschudi2021, benisty23}, which aligns closely with Dip \#1 and Dip \#2. We also note that this expected depolarization paired with possible coronagrphic residuals can lead to increased minor axis depolarization, as seen with the unsubtracted coronagraphic halo in \cite{Hashimoto2012} PDS 70 observations. 

\subsection{Degeneracies with the Scattering Background}\label{subsec:background} 
It should be noted that when interpreting azimuthal brightness variations as shadows, an inclined disk is also subject to azimuthal brightness variation caused by changing scattering angle \citep{Schmid2021}.
In the case of MWC 480, there is expected depolarization along the disk minor axis (PA \ang{53.6}) in the vicinity of the observed brightness dips. Thus we must disentangle the dips from this scattering background to determine true movement. Following \cite{Milli2019}, we attempt to analytically model the background scattering. A detailed breakdown of the procedure can be found in appendix~\ref{appendix MWC480}. Scattering can be defined using the Henyey-Greenstein (HG) model of anisotropic scattering \citep{Henyey1941}. The HG scattering parameter $g$ ranges from [-1, 1] and $g<0$ defines backwards scattering, $g=0$ is isotropic scattering, and $g>0$ is forward scattering. Unfortunately, the $g$ parameter for MWC 480 has no previous definition, and \cite{Garufi2014} constrained $g$ for HD 163296 to be $0<g<0.6$, leaving a significant range of potential $g$ values. This broad range of $g$ corresponds to an equally broad range of potential background azimuthal profiles which directly impact the apparent location of the brightness peaks (see Figures \ref{fig: testing g} and \ref{fig: testing g HD163296}). 
From this test we conclude that the level of background scattering cannot be constrained without radiative transfer modeling or without a defined scattering parameter, and thus we cannot provide a true measure of dip movement independent of background scattering. While variation in the surface brightness distribution is apparent, the exact location of the dips cannot be separated from the background and therefore limits our ability to speculate on the origin of these dips. 

Further scattered light observations to test if the HD 163296 Peak \#1 has continued to move in a consistent direction, as well as robust inner disk models will help illuminate what the cause of this feature may be. 

\subsection{Reason for Non-detections} \label{subsec:non-detections discussion}
We detect three disks with SCExAO/CHARIS at $>2\times$ lower SNR than with SPHERE/IRDIS (Figure~\ref{fig:SNR MWC480 AP}, appendix \ref{app: instrument}), however the angular resolution of the CHARIS observations is sufficient to recover the same disk features seen with IRDIS.
Our three successful disk detections indicate that the experimental set-up and data-reduction methods are capable of recovering disk signals at SNR$>$3. Therefore, the lack of disk detections may be linked to a difference in properties between the six non-detected targets and three detected ones. The stellar and disk properties for our nine targets are listed in Table~\ref{tab:observations}. We examined the parameters to search for a correlation between the three targets that were successfully detected and the ones that were not. We see no trend in stellar mass or luminosity correlating to disk detection. Only two of the six non-detections (PDS 76 and HD 144432) had disk mass estimates from mm flux available \citep{Guzman2021}, both of which are on the same order of magnitude as our detected disks, so there does not appear to be an obvious correlation between disk mass and detectability in our sample. We can note, however, that the three detected disks have the largest masses among those with available measurements. 

Of the six non-detections, four have available SEDs in \citet{Guzman2021} : HD 144432, HD 56895, PDS 76, and HD 81474 (note that the HD 56895 data has limited points, and any inferences from this SED should be taken with caution). 
Herbig SEDs are classified into groups I and II following the scheme of \citet{Meeus2001}. In this framework, group I SEDs can be represented by a combination of a power law and a blackbody component, while group II SEDs are described by a single power law. Group-I Herbig disks have large cavities, while group-II have less depleted cavities and are often assumed to be puffy in the inner disk and thus self-shadowed \citep{Garufi2022,Brittain2023}. This self-shadowing of group II disks has the possibility to impede a PDI detection. Only two out of six of our disks with a listed Meeus group are group I: HD 163296 and HD 143006, the two most prominently detected. The other five disks with Meeus listings are group-II which are often smaller in the IR and self-shadowed, and thus more difficult to detect. Therefore, this may potentially explain the lack of detection for the majority of our targets. Notably, the only group-II disk in our sample with a detection, MWC 480, was observed with a coronagraph, which may have been crucial in suppressing the stellar PSF, and allowing the disk to become resolvable. 

We note that PDS 76's disk has been detected in mm continuum \citep{Stapper2022}, and faintly in the NIR \citep{Garufi2017}, while HD 144432 has only faintly been detected in NIR (\cite{Perez2004}, \cite{Garufi2022}). The other disks in our sample have no NIR detections at this time. 

\section{Summary and Future Outlook}\label{sec:conclusions}
We conducted a near-infrared polarimetric imaging survey of nine Herbig systems selected from a volume-limited sample of Herbig disks within 200 pc \citep{Dong2018} with Subaru/SCExAO using both CHARIS and fast-PDI observing modes. We detected scattered-light disks in three targets (MWC 480, HD 163296, and HD 143006), while the remaining six systems were not detected. Our main results are summarized below. 

\begin{itemize}
    \item We detect two azimuthal brightness dips on the ring at $0\farcs19$ in the MWC~480 disk(Figure~\ref{fig: MWC 480 4}) which have not been previously highlighted (see section \ref{subsec: mwc480}).
    
    \item MWC 480 (Figure~\ref{fig: MWC profile}) and HD 163296 (Figure~\ref{fig:HD163296 All}) both show evidence of temporal variation in illumination. We see significant shifting of one of the brightness dips in MWC 480 (section \ref{subsec: mwc480}), and a potentially linear shift for the location of a brightness peak in HD 163296's ring at a deprojected distance of 0\farcs64 (section \ref{subsec:HD163296 analysis}, Figure~\ref{fig:HD163296 Move}). These changes are unlikely caused by instrumental polarization effects as those have been corrected. The changes do not correspond to local Keplerian movements of material given the low Keplerian velocities. We cannot determine the exact origin of these illumination changes without an understanding of the disk background scattering (section~\ref{subsec:background}, Appendix~\ref{appendix MWC480}). Additional observational epochs and inner disk modeling may provide more conclusive answers for their origin. 
    
    \item We do not detect any changes in shadow lane position, or peak brightness in HD 143006 (Figure~\ref{fig: HD143006 azimuthal}). 
    Our results remain consistent with the model of a tilted inner disk casting a shadow on the western side of the IR disk.
    
    \item We present the first published fast-PDI disk image of a protoplanetary disk (Figure~\ref{fig:fast_PDI}, Figure~\ref{fig:SNR HD143006 AP}). We resolve the brighter Ring \#2 in HD 143006 but fail to detect Ring \#1 due to insufficient SNR. We did not detect any new disks with this imaging mode. SCExAO/CHARIS has $>2\times$ smaller SNR than SPHERE/IRDIS at $K$-band (Figure~\ref{fig:SNR MWC480 AP}, appendix \ref{app: instrument}). However, CHARIS still successfully observed the three target disks with SNR$>$3, and sufficient resolution to detect the same disk features seen with IRDIS.
    
    \item No new Herbig disks were detected in HD~144432, HD~56895, PDS~76, HIP~80425, HD~148352, and HIP~81474 (Table~\ref{tab:observations}, Figure~\ref{fig:CHARIS_PDI}), possibly attributed to the outer disks being too small or self-shadowed (as understood by Meeus grouping) and thus overwhelmed by the PSF (section \ref{subsec: non detections pol signal}). For these targets, we report the wavelength dependence of the polarization fraction and angle of linear polarization (AoLP), including the significant polarization variation seen in HD 144432 (Figure~\ref{fig:AOLP and percent}, Table~\ref{tab: pol frac}).
\end{itemize}

Our current observations do not represent the full potential of SCExAO. The AO system has recently been upgraded to AO3k which introduces a 3000-actuator deformable mirror as the primary facility AO system, leading to an increase of 1000 actuators and improved wavefront control \citep{Lozi2024}. This should improve PDI sensitivity and provide a smaller inner working angle, both of which could be crucial to detect potentially faint disks such as the 6 non-detections in our sample and MWC 480. Future observations of these targets with the upgraded AO system could 
provide a comparable epoch to help diagnose the origin of each disk's brightness variations. 

{%
\let\internallinenumbers\relax
\nolinenumbers
\begin{acknowledgments}
We thank the anonymous referee for providing thoughtful and
constructive questions and suggestions. We acknowledge the contributions of Kyohoon Ahn, Vincent Deo, and S\'ebastien Vievard as members of the SCExAO instrumentation team, whose work in instrument and software design were critical for our data acquisition. We thank Nobuhiko Kusakabe, Bin Ren, Christain Ginski, and John Monnier for providing fits files.  We also thank Erika Dykes for assistance in calculating SNR for CHARIS targets and Thayne Currie for assistance with understanding our non-detections.
The development of SCExAO is supported by the Japan Society for the Promotion of Science (Grant-in-Aid for Research \#23340051, \#26220704, \#23103002, \#19H00703, \#19H00695 and \#21H04998), the Subaru Telescope, the National Astronomical Observatory of Japan, the Astrobiology Center of the National Institutes of Natural Sciences, Japan, the Mt Cuba Foundation and the Heising-Simons Foundation. The authors wish to recognize and acknowledge the very significant cultural role and reverence that the summit of Maunakea has always had within the indigenous Hawaiian community, and are most fortunate to have the opportunity to conduct observations from this mountain. CHARIS was built at Princeton University under a Grant-in-Aid for Scientific Research on Innovative Areas from MEXT of the Japanese government (\# 23103002). 
This work was supported by the National Science Foundation of China (32450631 and 62371007), JSPS KAKENHI Grant Number 23K03463, and the Natural Sciences and Engineering Research Council of Canada (NSERC; RGPIN-2023-05299).
C.M. is supported by the Natural Sciences and Engineering Research Council of Canada (NSERC) through a Canada Graduate Scholarships – Doctoral (CGS-D) award (CGS D - 600093 - 2025). 
D.J.\ is supported by NRC Canada and by an NSERC Discovery Grant.
This research was enabled in part by support provided by the Digital Research Alliance of Canada (\url{alliance.can.ca}), the National Energy Research Scientific Computing Center (NERSC) under award number FES-ERCAP-m4239, and by the High-performance Computing Platform of Peking University.
\end{acknowledgments}
}%
%

\appendix
\restartappendixnumbering
\section{Disk background Model} \label{appendix MWC480}
\begin{figure} [!htpb]
    \centering
    \includegraphics[width=0.9\linewidth]
    {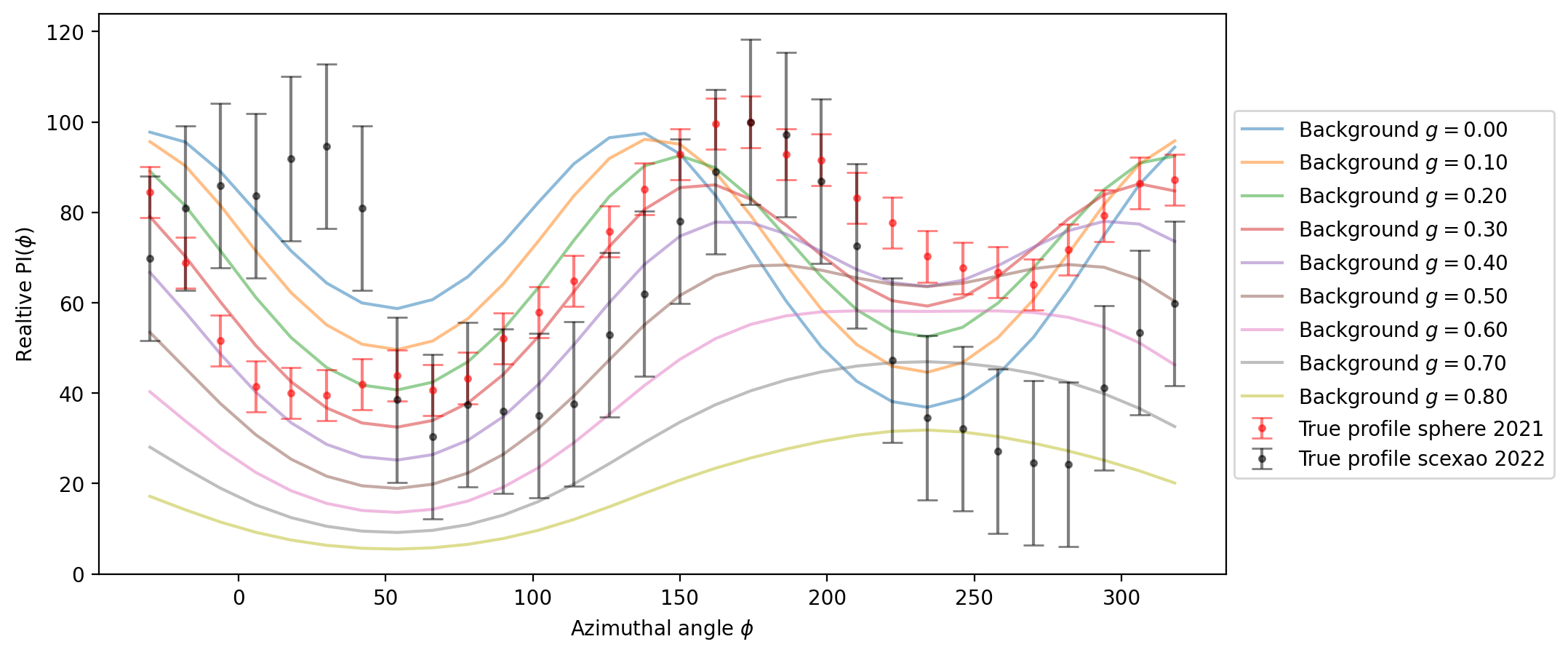}
    \caption{
    Background scattering model testing a range of $g$ parameters. The value of $g$ greatly impacts the degree of expected depolarization along the east and west minor axis; however the location of the depolarization dip centers are not affected.
    }
    \label{fig: testing g}
\end{figure}

\begin{figure} [ht]
    \centering
    \includegraphics[width=0.7\linewidth]{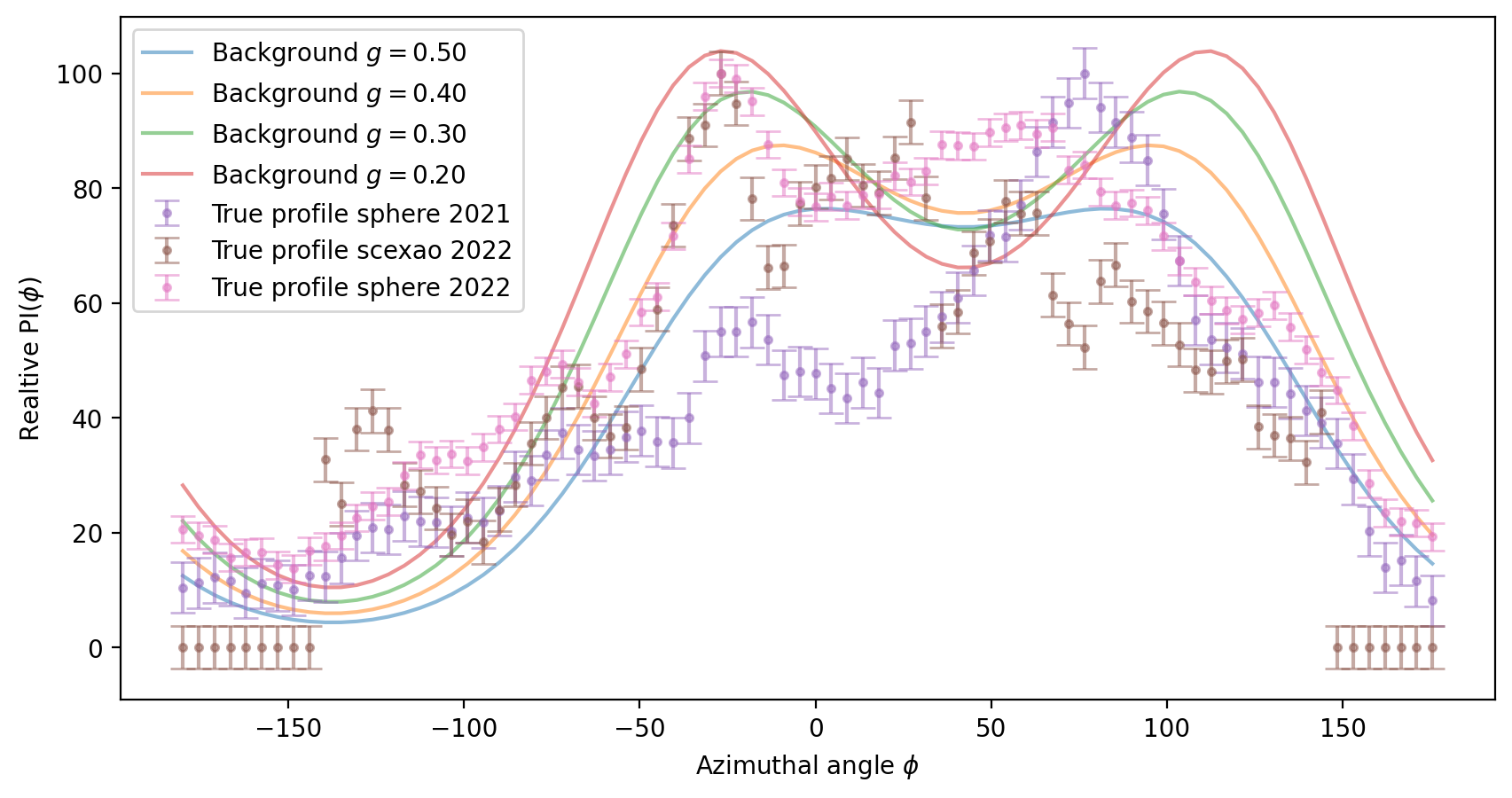}
    \caption{Background Scattering model testing a range of $g$ parameters for HD 163296 with the knowledge that $0<g<0.6$.}
    \label{fig: testing g HD163296} 
\end{figure}
Here we present background scattering modeling 
by combining the Rayleigh scattering fraction ($p$) for a single scatter model, with Henyey-Greenstein (HG, \citealt{Henyey1941}) model of anisotropic scattering. 

We begin by defining the scattering angle 
\begin{equation} \label{eq: scattering angle}
    \theta_s = \arccos(\sin i\cos \phi \cos \alpha + \sin \alpha\cos i)
\end{equation}
where $i$ is the disk inclination, $\alpha$ is the flaring angle, and $\phi$ is the  position angle from the forward-minor axis. 
The $\theta_s$ from equation~\ref{eq: scattering angle} is then applied to the Rayleigh 
degree of polarization
\begin{equation} \label{eq: Rayleigh}
    p = \frac{1-\cos^2 \theta_s}{1+\cos^2 \theta_s}
\end{equation}
and the HG function 
\begin{equation} \label{eq: HG}
    HG = \frac{1-g^2}{(1+g^2-2g\cos \theta_s)^\frac{3}{2}}
\end{equation}
where $g$ ranges from [-1, 1] and $g<0$ defines backwards scattering, $g=0$ is isotropic scattering, and $g>0$ is forward scattering. Following methods of \cite{Milli2019} and \cite{Schmid2021}, equations \ref{eq: Rayleigh} and \ref{eq: HG} are multiplied to get the intrinsic scattering of the disk dust in polarized NIR light. To account for instrumental effects, we add a FWHM gaussian blur ($Gb$) to obtain our final polarization function: 
\begin{equation}
    f(g, \theta_s) = p\times HG\times Gb 
\end{equation}

Nearly all the necessary parameters for the SCExAO observation can be directly calculated or estimated with MCMC, except the HG $g$ parameter which cannot be determined directly from a singular epoch which may posses shadowing. The $g$ value for NIR observations is not expected to exceed 0.85 \citep{Hughes2018}, however this still leaves a large range of potential $g$ values. We tested the model on a range of $g$ values,  shown in Figures~\ref{fig: testing g} and \ref{fig: testing g HD163296}. We find that it greatly impacts the expected depth of minor-axis depolarization and would consequently impact the positioning of the extracted brightness dip residual. In order to constrain the theoretically nearly static scattering background from potential dynamic shadowing, we would require several more observation epochs, such that an expected average background scattering in the absence of shadowing can be established. Since we only possess two useful epochs, we cannot disentangle the scattering background from dip positioning without a robust knowledge on $g$.

\section{Instrument Comparison} \label{app: instrument}
\begin{figure}[!htpb]
    \centering
    \includegraphics[width=0.99\linewidth]{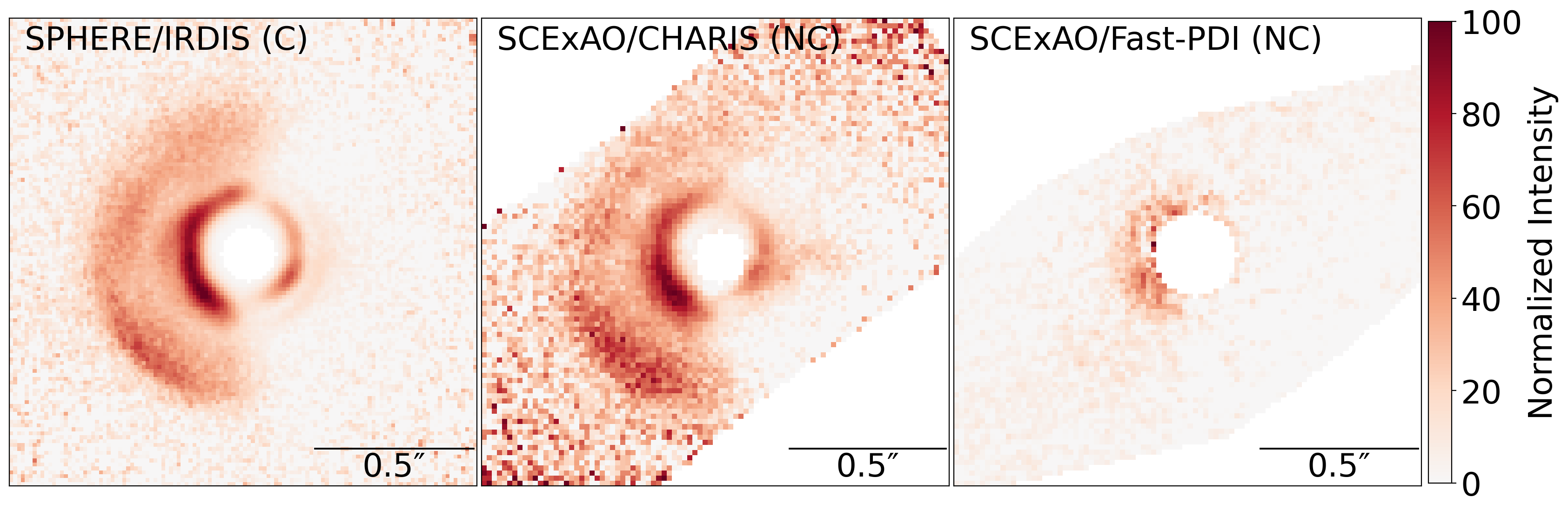}
    \caption{Visual instrumental comparison of SPHERE/IRDIS, SCExAO/CHAIRS and SCExAO/Fast-PDI featuring HD 143006. The ``C" label indicates coronagraphic observations while the ``NC" indicates non-coronagraphic.}
    \label{fig:SNR HD143006 AP}
\end{figure}
\begin{figure}[!htpb]
    \centering
    \includegraphics[width=0.95\linewidth]{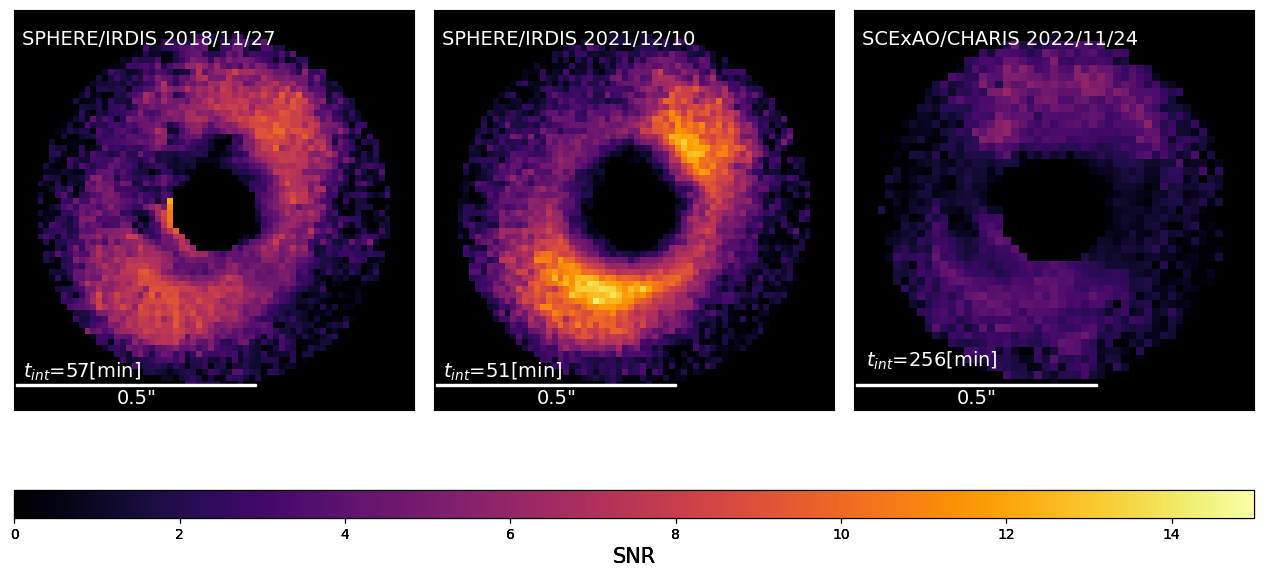}
    \caption{SNR for three epochs of MWC 480. All epochs are coronagraphic making for the best comparison among our available detections}.
    \label{fig:SNR MWC480 AP}
\end{figure}

SCExAO/CHARIS, SCExAO/fast-PDI, and SPHERE/IRDIS are compared with one another for HD 143006 in Figure~\ref{fig:SNR HD143006 AP}. While IRDIS shows higher resolution than CHARIS, the important disk features are also resolvable for CHARIS. By contrast, fast-PDI only reveals Ring\#2, but cannot resolve Ring\#1, nor the west side of Ring\#2 which allows the shadow lanes to become clear.

In Figure~\ref{fig:SNR MWC480 AP} we compare SNR for the only disk where we have coronagraphic observations -- MWC 480 -- and thus can most accurately compare with the SPHERE/IRDIS data. IRDIS achieves greater SNR than CHARIS, reaching a peak SNR of $\sim16$ as opposed to $\sim6$ for $\sim1/4$ the exposure time. Given that MWC 480 is our only detected Meeus group 2 disk, it is possible that it's relatively long exposure time was crucial in detecting the disk.

\section{Additional Figures and Tables} \label{appendix}
\begin{figure*}[htpb]%
    \centering
    \includegraphics[width=0.99\textwidth]{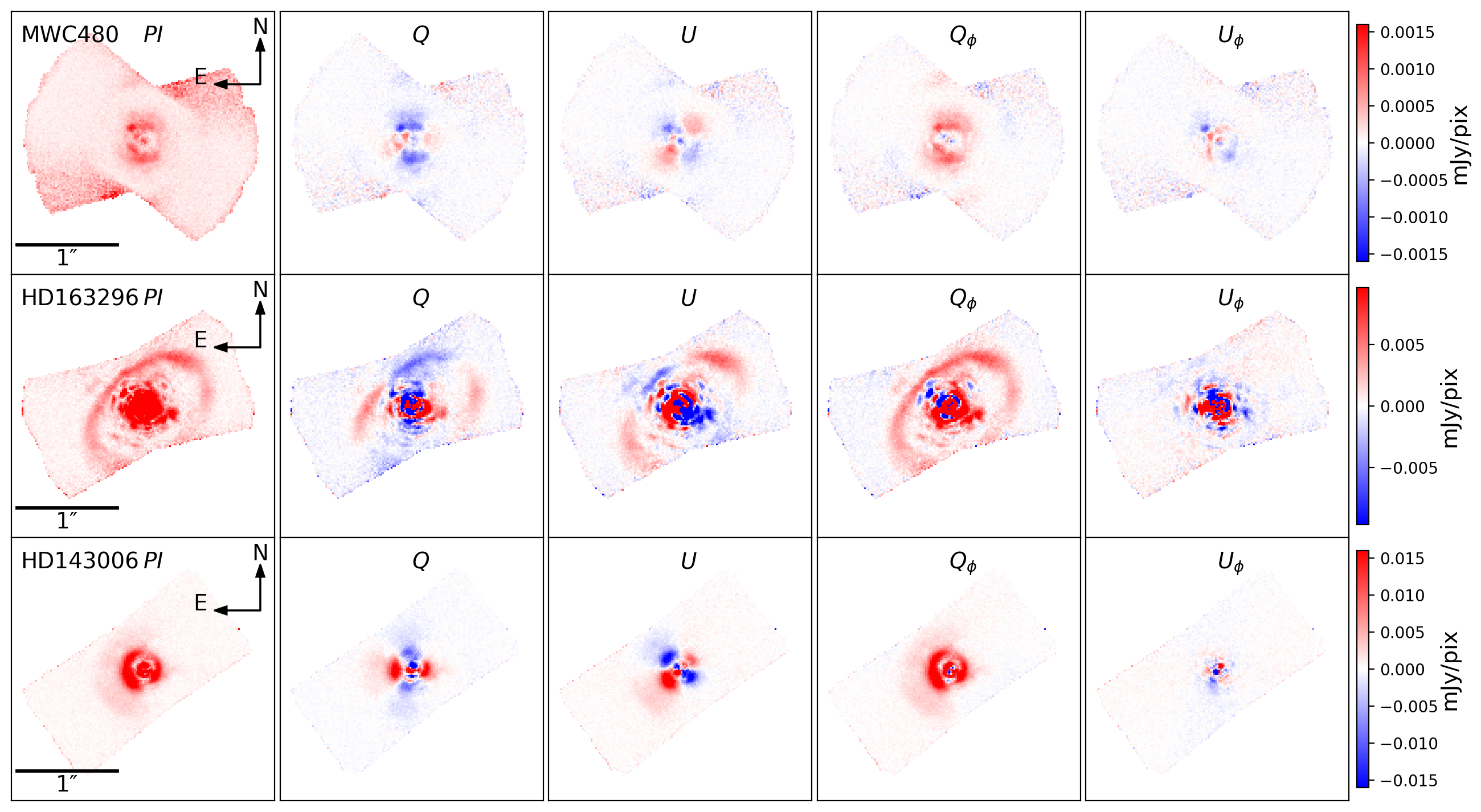}
    \caption{CHARIS PDI results showing $PI$, $Q$, $U$ 
    and azimuthal stokes $Q_\phi$ for the 3 previously-detected disks. Each disk is detected as can be seen by the positive $Q_\phi$ and noise-like $U_\phi$, as well as the characteristic quadrupole pattern in Q and U in the expected orientation. (The $PI$, $Q$, and $U$ data are available in the online version.)
    } 
    \label{fig:previously observed AP}
\end{figure*}

\begin{figure*}[!htpb]%
    \centering
    {\includegraphics[width=0.95\textwidth]{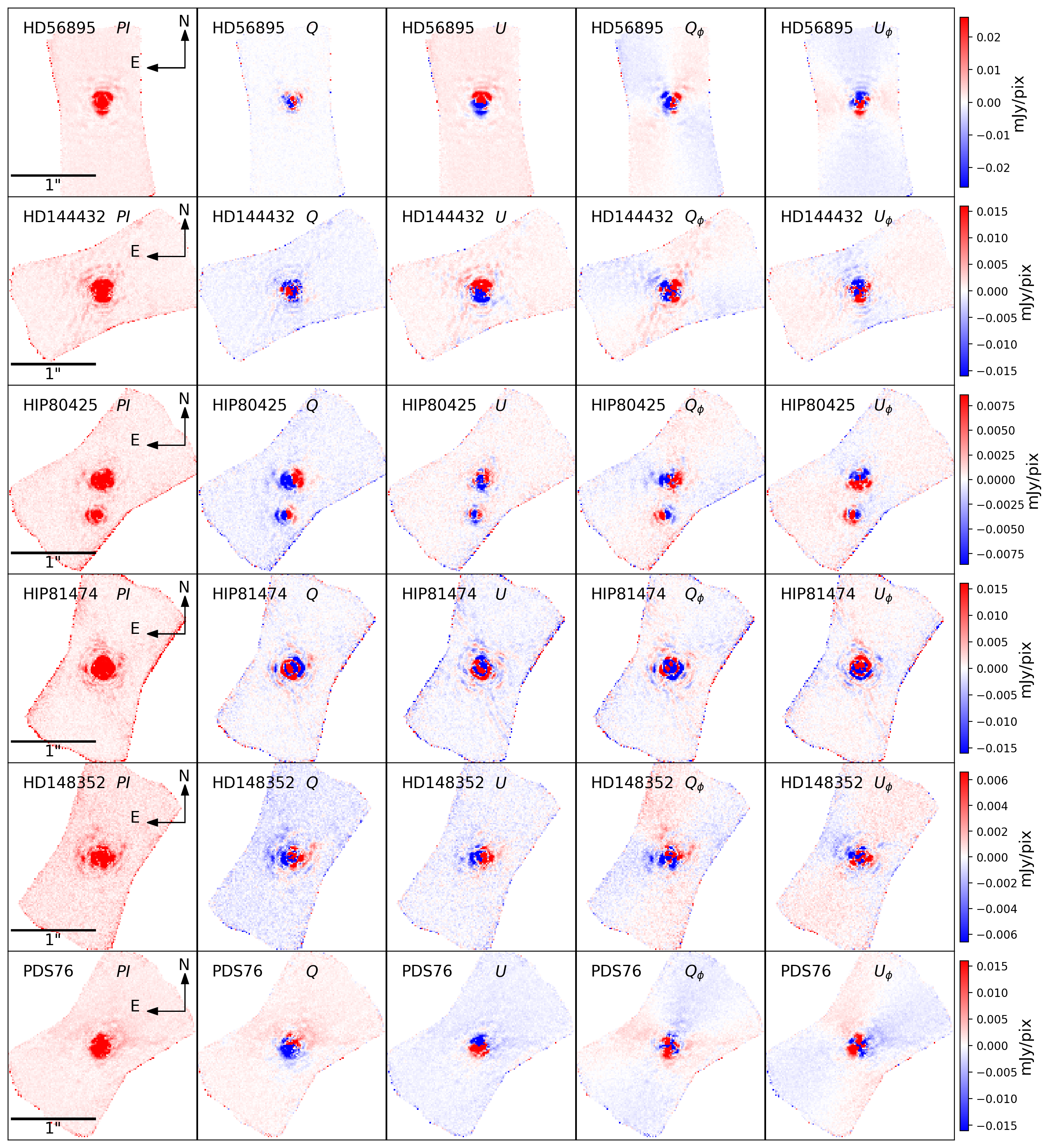}}
    \caption{
    CHARIS PDI results showing $PI$, $Q$, $U$ and azimuthal stokes for the six new targets. The color bar is set such that red is positive and blue is negative. (The $PI$, $Q$, and $U$ data are available in the online version.)}
    \label{fig:CHARIS_PDI AP}
\end{figure*}

\begin{figure*}[!htpb]%
    \centering
    \includegraphics[width=0.99\textwidth]{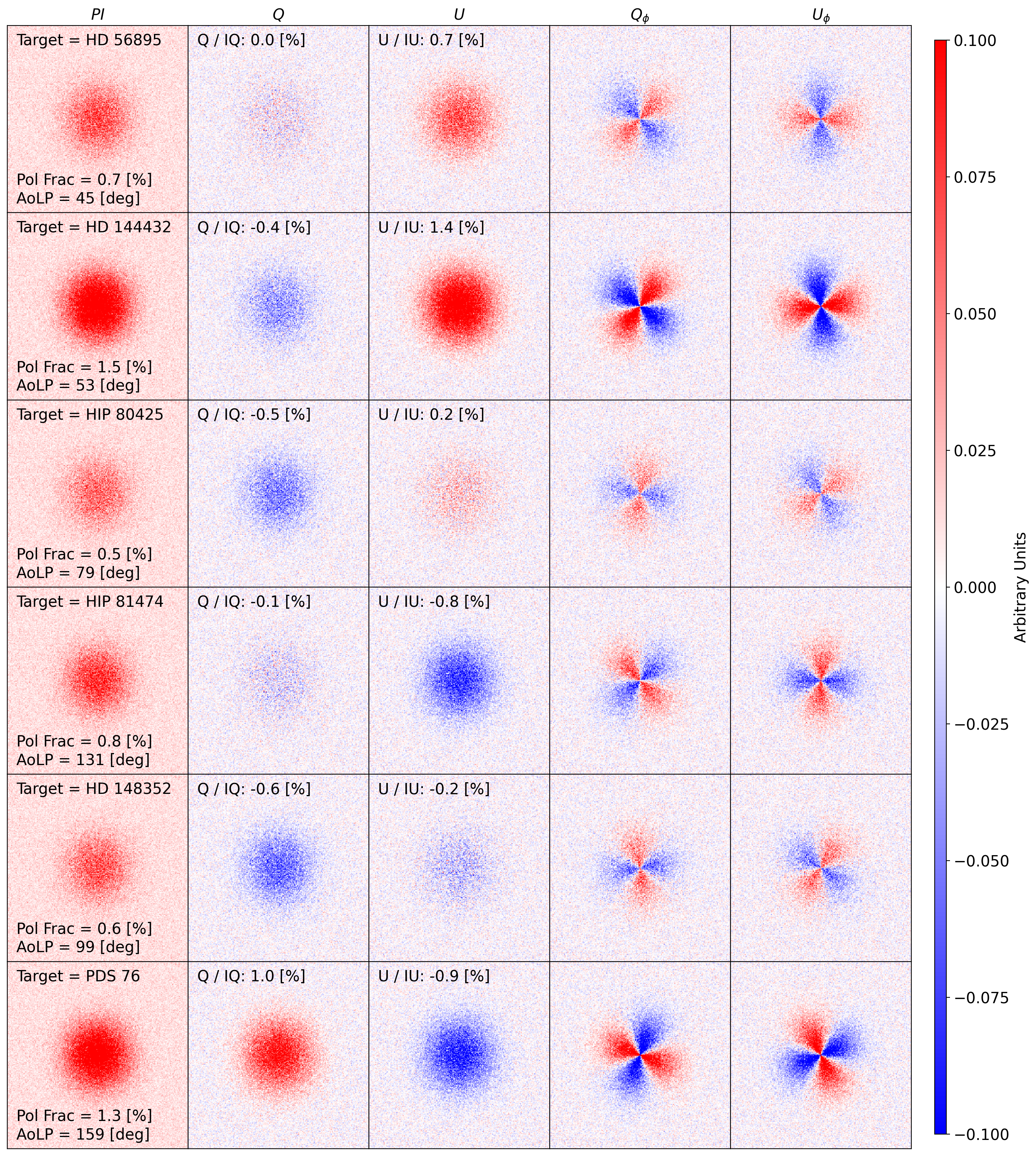}
    \caption{Each row demonstrates the theoretical double-differential-imaging $PI$, $Q$, $U$, $Q_\phi$ and $U_\phi$ products for a PSF with randomized noise which is polarized to a specific degree (Pol Frac) in a specific direction (AoLP). The polarization fraction and AoLP applied to each model are measured from the calibrated CHAIRS data products and averaged across the 17 $K$-band wavelength channels. For plotting purposes the fractions are scaled up by 10\% to allow for better visibility of the patterns. However, the listed fractions on each plot are the true fractions for each target model.The measured polarization fraction in each stokes vector are listed as Q/IQ and U/IU. 
    } 
    \label{fig:Theoretical Q U AP}
\end{figure*}

\begin{figure}[!htpb] 
    \centering
    \subfigure{\includegraphics[width=0.45\textwidth]{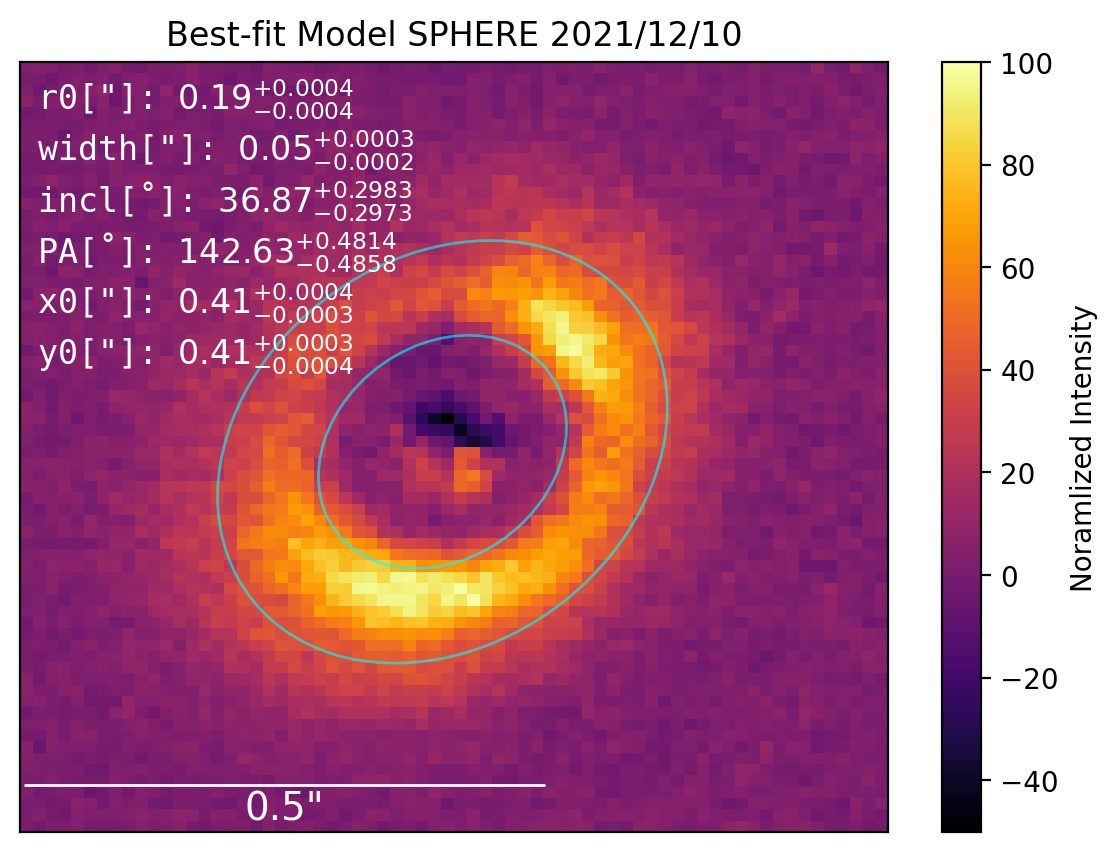}}
    \subfigure{\includegraphics[width=0.45\textwidth]{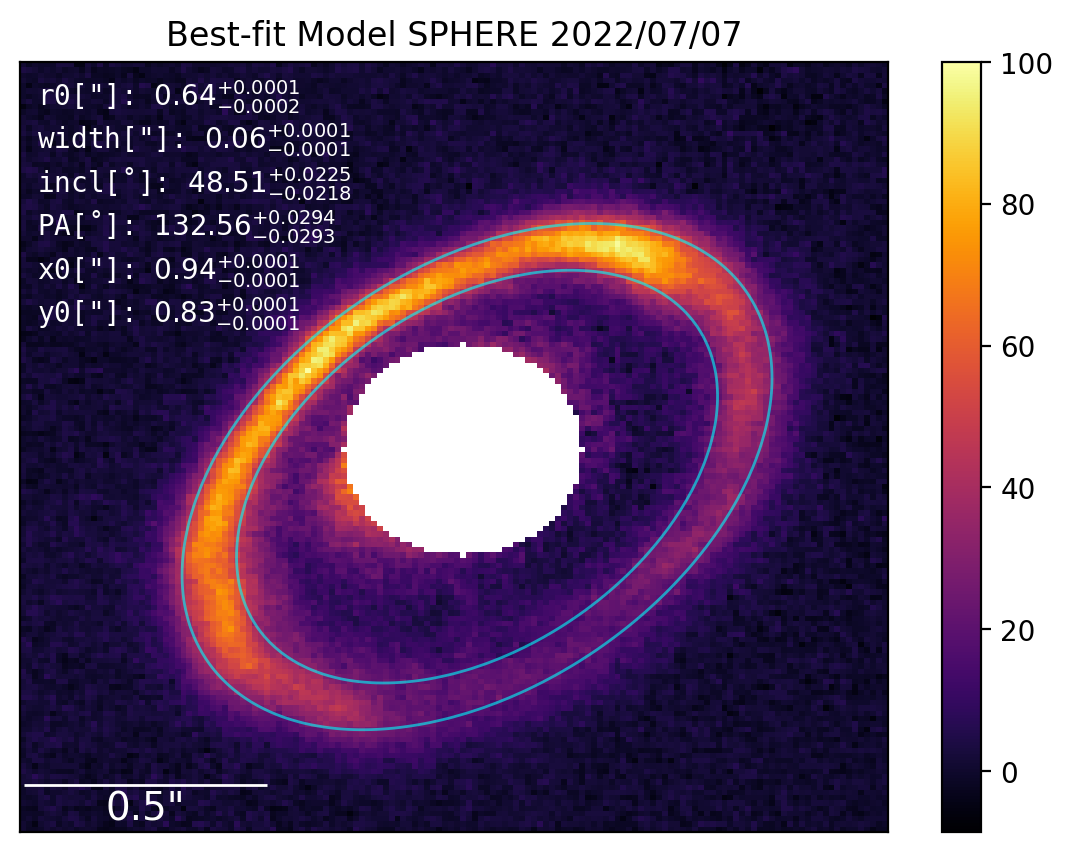}}
    \caption{
    \textbf{Left}: Fit of gaussian disk model using MCMC for MWC 480 SPHERE 2021 epoch. Best fit parameters with 1$\sigma$ errors are listed. The blue ellipses show the radial boundaries of the fitted ellipse. 
    \textbf{Right}: MCMC fit to HD 163296 SPHERE 2022/07/07 epoch. Best fit parameters with 1$\sigma$ errors are listed. The blue ellipses show the radial boundaries of the fitted ellipse. }
    \label{fig: ring fit MWC 480}
\end{figure}

\begin{figure*}[!htpb] 
    \centering
    \includegraphics[width=0.99\textwidth]{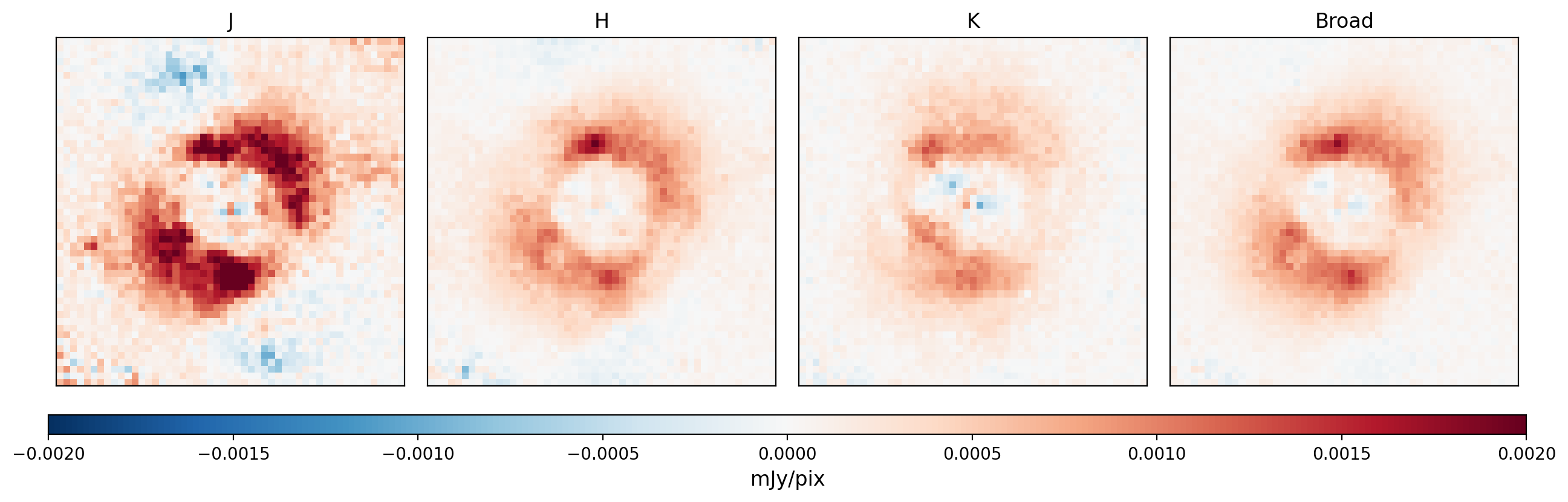}
    \caption{
    Median-combined SCExAO CHARIS $Q_\phi$ images of MWC 480 at  $J$, $H$, $K$-band, and the full $JHK$ broadband image. (The broadband fits files are available in the online version.)
    }
    \label{fig: color_compare_MWC480}
\end{figure*}

In this appendix we show additional figures which complement the main text. Figures \ref{fig:previously observed AP} and \ref{fig:CHARIS_PDI AP} showcase the full $PI$, $Q$, $U$ and azimuthal stokes for our detected and non-detected disks respectively. Figure \ref{fig:Theoretical Q U AP} showcases the theoretical point-source $PI$, $Q$, $U$ and azimuthal stokes outputs for different polarization percentages. 
Figure \ref{fig: ring fit MWC 480} shows the disk model boundaries defined by MCMC for MWC 480 and HD 163296.  Figure \ref{fig: color_compare_MWC480} shows additional wavelength images of MWC 480.


\bibliography{Mullin_SCExAO}{}
\bibliographystyle{aasjournal}



\end{document}